\documentclass[preprint,showpacs,preprintnumbers,amsmath,amssymb]{revtex4}
\usepackage{graphicx}
\usepackage{dcolumn}
\usepackage{bm}
\usepackage{amssymb}
\usepackage{amsmath}
\usepackage{latexsym}

\begin{document}

\title{From Dinuclear Systems to Close Binary Stars: Application to Source of Energy in the Universe}

\author{V.V. Sargsyan$^{1,2}$, H. Lenske$^{2}$, G.G. Adamian$^{1}$, and N.V. Antonenko$^{1}$}

\affiliation{
$^{1}$Joint Institute for Nuclear Research, 141980 Dubna, Russia,\\
$^{2}$Institut f\"ur Theoretische Physik der
Justus--Liebig--Universit\"at, D--35392 Giessen, Germany}

 \date{\today}

\begin{abstract}
The evolution of  close binary  stars in mass asymmetry (transfer) coordinate is considered.
The conditions for the formation of stable symmetric binary stars are analyzed.
The role of   symmetrization  of asymmetric binary star in the transformation
of  potential energy into internal energy of star and the release of a
large amount of energy is revealed.
\end{abstract}

\pacs{26.90.+n, 95.30.-k \\ Keywords:
close binary stars, mass transfer, mass asymmetry}

\maketitle

\section{Introduction}\label{sec:Intro}
Because mass transfer is an important observable for close
binary systems in which the two stars are nearly in contact \cite{Kopal:1978,Shore:1994,Hild:2001,Boya:2002,Eggleton:2006,Khal:2004,Vasil:2012,Cher:2013},
it is meaningful and necessary to study the evolution of these stellar
systems in the mass asymmetry coordinate
$\eta = (M_1-M_2)/(M_1+M_2)$ where $M_i$, $(i =1, 2)$, are the stellar
masses.
In our previous work \cite{IJMPE}, we used
classical
Newtonian mechanics and study the evolution of the close binary stars in
their center-of-mass frame by analyzing the
total potential energy $U(\eta)$ as a function of $\eta$ at fixed total mass $M=M_1+M_2$
and orbital angular momentum $L=L_i$ of the system. The limits for the
formation and evolution of the di-star systems were derived and analyzed.
We used theoretical methods which are being applied successfully to
corresponding processes in nuclear systems where mass asymmetry plays an
important role as the collective coordinate governing fusion of two heavy nuclei
\cite{Adamian:2012,Adamian:2014}. Nuclear dynamics, of course, is quite different
from the gravitational interactions in di-stars. Nuclear reactions are
dominated by short-ranged strong interactions, to which minor
contributions of long-range (repulsive) Coulomb and centrifugal forces are superimposed.
However,  extending the methods and results from the femto-scale of
microscopic nuclear physics   to macroscopic binary stellar
systems, we obtained that  the driving potentials $U(\eta)$ for the di-star systems looks
like the driving potentials for the microscopic dinuclear systems \cite{IJMPE,Adamian:2012,Adamian:2014}.

One should stress that in Ref. \cite{IJMPE} the mass-asymmetry-independent structural factor
was used as a free parameter.  In the general, the dimensionless structural factor $\omega_i$
is   determined by   the density profile  of the star.  In the present paper, we generalize the approach of Ref. \cite{IJMPE} by
employing   the values of the structural factor of the stars from  the model of Ref. \cite{Vasil:2012},
Note that the model of Ref. \cite{Vasil:2012} well describes the
observable temperature-radius-mass-luminosity relations of stars, especially binary stars,
the spectra of seismic oscillations of the Sun, distribution of stars on their masses, magnetic
fields of stars and etc. The stellar radii, masses and temperatures are expressed by the corresponding ratios of the fundamental
constants, and individuality of stars are determined by two parameters - by the charge and mass numbers of nuclei,
from which a star is composed \cite{Vasil:2012}.

Our theoretical approach is introduced in Sect. \ref{sec:Theory}.
We explore the landscape of the total potential
energy of the close binary system, searching
specifically for the evolution paths in the mass asymmetry coordinate. In Sect.
\ref{sec:Results} the theoretical method is applied to concrete close
binaries. It is shown that the mass transfer is one of the important
sources of the transformation of the gravitational  energy
to other types of energy in the universe. In Sect. \ref{sec:Summary}
 the obtained results are   summarized.

\section{Theoretical Method}\label{sec:Theory}

The differential of total energy of di-star system as a function of relative distance ${\bf R}$ of two stars,
conjugate canonical momentum ${\bf P}$ and mass asymmetry coordinate $\eta$ reads as
\begin{eqnarray}
dE({\bf R},{\bf P},\eta)=\frac{\partial E}{\partial t}dt+\frac{\partial E}{\partial {\bf R}}d{\bf R}+\frac{\partial E}{\partial {\bf P}}d{\bf P}+\frac{\partial E}{\partial \eta}d\eta.
\label{dEtot}
\end{eqnarray}
The kinetic energy in $\eta$ is assumed to be small and can be disregarded.
In the center of mass system, the total energy of the di-star system is a sum of radial and orbital part of kinetic energies, and the potential energy.
For the reasons seen below, we attach the orbital kinetic energy part to the star-star interaction $V$.
In this case the expression of the total energy of the di-star system reads as
\begin{eqnarray}
E=\frac{P_{R}^2}{2 \mu}+U,
\label{Etot}
\end{eqnarray}
where $P_{R}$ is the radial component of the momentum ${\bf P}$ and
$\mu=\mu(\eta)=\frac{M_1M_2}{M}=\frac{M}{4}(1-\eta^2)$ is the reduced mass.
The total potential energy of the di-star system
\begin{eqnarray}
U=U_1+U_2+V
\label{eq_pot}
\end{eqnarray}
is given by the sum of the potential energies $U_i$ ($i=1,2$) of the two stars and star-star interaction potential $V$.
The radiation energy is neglected because
 the absolute values of the gravitational energy and the intrinsic kinetic energy
 are much  larger than  the radiation energy.
The energy of the star  "$i$" is
\begin{eqnarray}
U_{i}=-\omega_i\frac{G M_i^2}{2R_i},
\label{eq_pot2}
\end{eqnarray}
where $G$, $M_i$, and $R_i$ are the gravitational constant, mass, and radius of the star, respectively.
Employing  the values of the dimensionless structural factor
\begin{eqnarray}
\omega_i=1.644\left(\frac{M_{\odot}}{M_i}\right)^{1/4}
\label{eq_omega2}
\end{eqnarray}
and radius
$$R_i=R_{\odot}\left(\frac{M_i}{M_{\odot}}\right)^{2/3}$$
of the star from  the model of Ref. \cite{Vasil:2012},
we obtain
\begin{eqnarray}
U_{i}&=&-\omega_0G M_{i}^{13/12}/2,\nonumber \\
\omega_0&=&1.644\frac{M_{\odot}^{11/12}}{R_{\odot}},
\label{eq_pot1}
\end{eqnarray}
where,   $M_{\odot}$, and $R_{\odot}$ are   mass  and radius of the Sun, respectively.
Because the average density $\rho_i=M_{\odot}\rho_{\odot}/M_i$ ($\rho_{\odot}$
is the average density of the Sun) increase with decreasing the mass $M_i$ of star \cite{Vasil:2012},
the structural factor $\omega_i$ depends on $M_i$ in Eq. (\ref{eq_omega2}).
The change of $\eta$ from 0 to 1 leads to the change  of $\omega_1$ by about of 16\%.
Note that in our work \cite{IJMPE}
the   structural factor $\omega_i$ was used as a free mass-asymmetry-independent  parameter.

Because the two stars rotate with respect to each other around the common center of mass,
the star-star interaction potential contains, together with the gravitational energy $Q$
of the interaction of two stars,  the kinetic energy of orbital rotation $V_{rot}$:
\begin{eqnarray}
V(R)=Q+V_{rot}=-\frac{GM_1M_2}{R}+\frac{\mu v^2}{2},
\label{eq_pot3}
\end{eqnarray}
where $v=(GM[2/R-1/R_m])^{1/2}$ and $R_m$ are the speed
and   the semimajor axis of  the elliptical relative orbit,
respectively \cite{Kopal:1978,Shore:1994,Hild:2001,Boya:2002,Eggleton:2006,Vasil:2012,Cher:2013}.
Finally, one can derive the simple expression for the star-star interaction potential
\begin{eqnarray}
V(R_m)=-\frac{GM_1M_2}{2R_m}.
\label{eq_pot5nn}
\end{eqnarray}
Because of Kepler's laws
\begin{eqnarray}
R_m=\left(\frac{\mu_i}{\mu}\right)^2R_{m,i},
\label{eq_pot5nnD}
\end{eqnarray}
where the index "i" denotes the initial value of reduced mass and distance between the initial (before transfer) binary star,
Eq. (\ref{eq_pot5nn}) is rewritten as
\begin{eqnarray}
V(R_m)=-\omega_VG(M_1M_2)^3/2,
\label{eq_pot5nnn}
\end{eqnarray}
where
$$\omega_V=\frac{1}{M^2\mu_i^2R_{m,i}}.$$

The final expression for the total potential energy (\ref{eq_pot}) of the di-star system is
\begin{eqnarray}
U=-\frac{G}{2}\left(\omega_0[M_1^{13/12} + M_2^{13/12}]+\omega_V[M_1M_2]^{3}\right).
\label{eq_pot6}
\end{eqnarray}
Using the mass asymmetry coordinate $\eta$ instead of masses $M_1=\frac{M}{2}(1+\eta)$ and $M_2=\frac{M}{2}(1-\eta)$,
we rewrite   Eq. (\ref{eq_pot6}):
\begin{eqnarray}
U&=&-\frac{GM_{\odot}^{2}}{2R_{\odot}}\left(\alpha[(1+\eta)^{13/12}+(1-\eta)^{13/12}]+\beta [1-\eta^2]^{3}\right),
\label{eq_pot7}
\end{eqnarray}
where
$$\alpha=1.644\left(\frac{M}{2M_{\odot}}\right)^{13/12}$$
and
$$\beta=\left(\frac{\pi^{2}M_{\odot}^{5}R_{\odot}^3}{32G\mu_i^6P_{orb,i}^{2}}\right)^{1/3}\left(\frac{M}{2M_{\odot}}\right)^{11/3}.$$
To obtain $\beta$, we use the Kepler's third law connecting
 the semimajor axis
\begin{eqnarray}
R_{m,i}= \left(\frac{ G M P_{orb,i}^2}{4\pi^2}\right)^{1/3}
\label{eq_pot4}
\end{eqnarray}
with the period $P_{orb,i}$ of  orbital rotation of the initial  di-star system
(orbital rotation of one star relative to the other).
As seen from Eq. (\ref{eq_pot7}), the stability of the binary star system depends on
the period  $P_{orb,i}$ or the orbital angular momentum
(with the mass asymmetry $\eta_i$) and  the total mass $M$.

Employing Eq. (\ref{eq_pot7}), we can study the evolution of the di-star system in the mass asymmetry coordinate $\eta$.
The extremal points of the potential energy as a function of $\eta$ are found by solving numerically the equation
\begin{eqnarray}
\frac{\partial U}{\partial\eta}=-\frac{GM_{\odot}^{2}}{2R_{\odot}}\left(\frac{13}{12}\alpha[(1+\eta)^{1/12}-(1-\eta)^{1/12}]-6\beta\eta[1-\eta^2]^{2}\right).
\label{eq_pot9}
\end{eqnarray}
As seen, Eq. (\ref{eq_pot9}) is solved for $\eta=\eta_m=0$.
At this value the potential  has an extremum which is a minimum if
$$\alpha  <\frac{432}{13}\beta$$
or
$$P_{orb,i}<\frac{128.5\pi}{(1-\eta_i^2)^3} \left(\frac{R_{\odot}^{3}}{GM_{\odot}}\right)^{1/2}\left(\frac{M}{2M_{\odot}}\right)^{7/8}$$
and a maximum if
$$\alpha  >\frac{432}{13}\beta.$$
The transition point is
$$\alpha_{cr}=\alpha=\frac{432}{13}\beta=\frac{432}{13}\left(\frac{\pi^{2}M_{\odot}^{5}R_{\odot}^3}{32G\mu_i^6P_{orb,i}^{2}}\right)^{1/3}\left(\frac{M}{2M_{\odot}}\right)^{11/3}.$$
If there is a minimum at $\eta=0$ ($\alpha<\alpha_{cr}$), it is engulfed symmetrically by two barriers.
Expanding Eq. (\ref{eq_pot9}) up to the third order in $\eta$ and solving it, we obtain the position of these barriers at $\eta=\pm\eta_b$, where
$$\eta_b=2^{-1/2}\left(\frac{864^2\beta-22464\alpha}{864^2\beta+3289\alpha}\right)^{1/2}.$$
So, at $\alpha<\alpha_{cr}$ the potential energy as a function of $\eta$ has two symmetric maxima at $\eta=\pm\eta_b$ and the minimum at
$\eta=\eta_m=0$. The fusion of two stars with $|\eta_i|<\eta_b$ can occur only by overcoming the barrier   at  $\eta=+\eta_b $ or $\eta=-\eta_b $.
With decreasing ratio $\alpha/\beta$,  $B_\eta=U(\eta_b)-U(\eta_i)$ increases and
the symmetric di-star system becomes more stable.
The evolution of two stars with $|\eta_i|\neq 0$ and $|\eta_i|<\eta_b$ to the symmetric di-star configuration is energetically favorable.
Hence, an initially asymmetric binary system ( $|\eta|=|\eta_i|<\eta_b$) is
driven to mass symmetry, implying a flow of mass towards equilibrium and
 increase of the internal energy of stars by the amount $\Delta U=U(\eta_i)-U(\eta=0)$ (Fig. 1a).
At $\alpha\ge\alpha_{cr}$, $\eta_m=\eta_b=0$ and we have the inverse $U$-type potential  with maximum at $\eta=0$.
In such a system, the fusion of stars (one star "swallows" the other star) is the
only mode of evolution in $\eta$ transforming in the end  the di-star into a mono-star
with the release of   energy $E_f=U(\eta_i)-U(\eta=1)$ (Fig. 1b).
%


As seen,
$\eta_b\to 2^{-1/2}\approx 0.71$  if $\beta\gg\frac{1}{66}\alpha$.
In this case the condition $0<\eta_b<2^{-1/2}$
 means that in the asymmetric system
 with   mass ratio   $M_1/M_2 > (1+2^{1/2})^2\approx 6$ the stars fuse.
 Thus, di-stars  with $|\eta| > \eta_b$ are
unlikely to exist for sufficiently long time. Indeed, the binary stars with a large mass
ratio are very rare objects in the universe.


\section{Application to Close Binaries}\label{sec:Results}

In the calculations, we assume that  the orbital angular momentum $L_i$ and the total mass $M$ are conserved during
the conservative evolution of the di-star in   mass asymmetry coordinate $\eta$.
The orbital angular momentum $L_i$ is calculated by using the experimental masses $M_i$
of stars and  period $P_{orb,i}$ of their orbital rotation (see Appendix A).

One can express the   potential energy (\ref{eq_pot7}) in   units of
$$u=u_1+u_2+xv=(1+\eta)^{13/12}+(1-\eta)^{13/12} + x (1-\eta^2)^{3},$$
where  $x=\beta/\alpha=\frac{GM_{\odot}^3R_{\odot}}{3.288L_i^2}\left(\frac{M}{2M_{\odot}}\right)^{47/12}$.
  As seen from Fig. 2, at relatively large $x > 0.025$
the $u$ as a function of $\eta$ has two asymmetric maxima  and the minimum at
symmetry.
The  decrease of $x$
leads to the change of the shape of potential energy in $\eta$: $\eta_b$ approaches $\eta_m=0$
and  the height of barrier $B_\eta$ in $\eta$ decreases.
At $x$=0.025,  we have the inverse $U$-type symmetric potential  with maximum at $\eta=0$, and
the di-star system becomes unstable with respect to the matter transfer coordinate.
The asymmetrization of system  is energetically favorable.
The parameter $x\sim M^{47/12}/L_i^2$ depends on $M$ and $L_i$. For example, one can
decrease $x$ with   increasing $L_i$ or decreasing $M$.

Various  di-stars have different $M$ and $L_i$ and, correspondingly, the potential energy shapes.
The  potential energies (driving potentials) $U(\eta)$ of the close di-star systems
versus $\eta$  are presented in Figs.~3--7.
 %
%
For all systems shown,
$\alpha< \alpha_{cr}$ or
$$L_i<[10.1GR_{\odot}M_{\odot}^3]^{1/2}\left(\frac{M}{2M_{\odot}}\right)^{47/24},$$
respectively the potential energies have
symmetric barriers at $\eta=\pm\eta_b$ and the minimum at $\eta=\eta_m=0$.
As seen in Fig. 3, the barrier in $\eta$ appears as a result of the interplay
between the total gravitational energy $U_1+U_2$ of the stars and the star-star interaction potential $V$.
Both energies have different behavior as a function of mass asymmetry: $U_1+U_2$ decreases and $V$ increases
with changing $\eta$ from $\eta=0$ to $\eta=\pm 1$.
One should stress that the driving potentials $U(\eta)$ for the di-star systems looks
like the driving potentials for the microscopic dinuclear systems \cite{Adamian:2012,Adamian:2014}.
Note that the same conclusion was derived in Ref. \cite{IJMPE} by using the mass-asymmetry-independent
(constant) structural  factors $\omega_i$.

%

The evolution of the di-star system   depends on the initial  mass asymmetry $\eta=\eta_i$ at its formation.
If the original di-star is asymmetric, but $|\eta_i|<\eta_b$, then it is energetically favorable to evolve
in  $\eta$ to a configuration in the global minimum at $\eta=0$, that is, to form a symmetric  di-star system.
The matter of a heavy star can move to an adjacent light star enforcing the symmetrization of di-star without additional driving energy.
The symmetrization  of asymmetric binary star leads to the decrease of potential energy $U$  or
the transformation of the potential energy into internal energy of stars.
The resulting symmetric di-star is created at large excitation energy.
For example, for the binary systems
RR Cen ($\eta_i$=0.65),
V402 Aur ($\eta_i$=0.66),
and
V921 Her ($\eta_i$=0.61),
the internal energies of stars
increase during symmetrization
by amount
$\Delta U=U(\eta_i)-U(\eta=0)=$
$2\times 10^{41}$,
$10^{41}$,
and
$10^{41}$ J,
respectively.
Because the most of close binary stars are asymmetric ones, the symmetrization process leads to the release of a
large amount of energy in these systems and can be an important source of
energy in the universe (see Tables I, II, and III).
Note that accounting for the loss of angular momentum
will lead to an increase of the value $\Delta U$.

If   $|\eta_i|>\eta_b$ or $\eta_b=0$,  the di-star system is  unstable and evolves towards the mono-star system, thus,
enforcing the asymmetrization of the di-star.
The matter is transferred from the light star  to the heavy star even without additional external energy.
We found only one  close binary system
$\alpha$ Cr B ($M_1=2.58M_{\odot}$, $M_2=0.92M_{\odot}$, $\omega_1=1.30$, $\omega_2=1.68$, $\beta/\alpha=0.039$),
for which  $|\eta_i|=0.47>\eta_b=0.33$ (Fig. 3).

Because the fusion barriers $B_{\eta}$ in $\eta$ are quite large for the  systems with $|\eta_i|<\eta_b$
 in Tables I, II, and III,
the formation of a mono-star from the   di-star system by the thermal diffusion in mass asymmetry coordinate
is strongly suppressed.
%
Perhaps,  the existence of barrier in $\eta$ is the reason why very asymmetric close double-stars
 systems  with   $|\eta_i|>\eta_b$   are rarely observed.
This imposes restrictions on the asymmetric configurations with $|\eta|>\eta_b$
of the di-star systems. There can not be a stable di-star
with a very light star, with only a fraction of  the mass of the heavy companion.

The value of  $\alpha$ becomes larger than  $\alpha_{cr}$ and the minimum in $U(\eta)$ disappears, and
di-star asymmetrization (fusion in $\eta$ coordinate) occurs as the result of
a release of matter from one of the stars or
an increase of orbital momentum due to the strong external perturbation, i.e. by the third object,
or  the spin-orbital coupling  in the di-star.
Because the close binary stars BW Aqr ($\eta_i=0.04$, $\eta_b=0.34$), V 889 Aql ($\eta_i=0.04$, $\eta_b=0.30$),
V 541 Cyg ($\eta_i=0.02$, $\eta_b=0.19$),  V 1143 Cyg ($\eta_i=0.02$, $\eta_b=0.27$), and IT  Cas ($\eta_i=0$, $\eta_b=0.49$)
have the smallest fusion barriers $B_{\eta}$ (Table I, Figs. 4 and 5), these systems may be a good candidates for such kind processes.

A spectacular recent case is KIC 9832227 which
was predicted \cite{Molnar:2017} to be   merge in 2022, enlightening the sky as a red nova.
For the fate of KIC 9832227 ($\eta_i=0.63$, $\eta_b=0.84$), we predict that a fast merger is excluded (see Fig. 7). This di-star
is driven instead towards the mass symmetry. The mass is transferring from heavy star to light one and the relative distance between
two stars and the period of the orbital rotation are decreasing.
A huge amount of energy $\Delta U\approx 10^{41}$ J is released during the symmetrization.
As seen in Fig. 7,   the di-stars  KIC 9832227 and RR Cen ($\eta_i=0.65$, $\eta_b=0.85$) have almost the same $\eta_i$, $\eta_b$, and
potential energy shapes. So, the observation of the RR Cen di-star is also desirable.
It should be stressed that the observational data of Ref. \cite{Socia}
negate the 2022 red nova merger prediction \cite{Molnar:2017}.

\section{Summary}\label{sec:Summary}

The  isolated  close binary star systems  evolve along well
defined trajectories in classical phase space.
We have shown that energy conservation is enough to fix the
trajectory of the system in the potential energy landscape defined
by the  total mass and orbital angular momentum of system.
Exploiting the stationarity of the total energy,
stability conditions were derived and investigated
as functions of the mass asymmetry parameter $\eta$.
We have shown that this collective degree of freedom plays
a comparable important role   in macroscopic object as well
as in microscopic dinuclear systems. In close di-star systems,
the   mass asymmetry coordinate
can govern the fusion (merger) and symmetrization (due to the mass transfer)
processes of two stars.
An interesting aspect is that once $\eta$ has been determined e.g.
by observation, it allows to conclude on the stellar
structure parameters $\alpha$ or, likewise, $\omega_i$.

Our novel theoretical treatment of di-star system is based
on the fact that after the formation process the lifetime of a di-star system is long
enough to reach equilibrium conditions in the mass asymmetry coordinate. Hence, we could conclude that
the system will be a member of the sample of all di-star and mono-star configurations
with the probabilities depending on the potential energies of a given configuration.
For all systems considered, $\alpha < \alpha_{cr}$ and the potential energies have
symmetric barriers at $\eta=\pm\eta_b$ and the minimum at $\eta=\eta_m=0$.
At  $\alpha < \alpha_{cr}$, two distinct evolution scenario arise.
Let the di-star system be  initially  formed with $\eta=\eta_i$.
The two stars start to exchange matter where the fate of the binary
depends critically on the mass ratio: If $|\eta_i| < \eta_b$ the system
is driven to the symmetric di-star configuration (towards a global minimum of the potential landscape).
However, if $|\eta_i| > \eta_b$ the system  evolves towards
the mono-star system.  All asymmetric close binary stars considered,
except $\alpha$ Cr B, satisfy the condition $|\eta_i| < \eta_b$
and in these systems the symmetrization process occurs.
Note that for many systems  $U(\eta=0)<U(|\eta|=1$.


In the case of $|\eta_i| < \eta_b$  ($\alpha < \alpha_{cr}$), the symmetrization of stars leads to the release of a
large amount of energy about $10^{41}$~J as radiation outburst and  matter   release.
Thus, the symmetrization of stars in close binary systems due to the mass transfer is one of the important
sources of the transformation of the gravitational  energy  to other types of energy in the universe.
The symmetrization of binary system will lead to $M_1/M_2\to 1$, $T_1/T_2\to 1$ ($T_i$
are the temperatures of   stars), $L_1/L_2\to 1$ ($L_i$ are the luminosities of   stars), $R_1/R_2\to 1$
which are observable quantities.


\section{Acknowledgements}
V.V.S.   acknowledge  the partial
supports from the Alexander von Humboldt-Stiftung (Bonn).
This work was partially supported by  Russian Foundation for Basic Research (Moscow)  and
DFG (Bonn).

\newpage

\appendix

\section{The star-star interaction potential}
Because the two stars rotate with respect to each other around the common center of mass,
the star-star interaction potential contains, together with the gravitational energy
of  interaction of two stars,  the kinetic energy of orbital rotation:
\begin{eqnarray}
V(R)=-\frac{GM_1M_2}{R}+\frac{L^2}{2\mu R^2},
\label{eq_pot3}
\end{eqnarray}
where $L$ is the orbital angular momentum of the di-star which is conserved during the conservative mass transfer.
From the conditions $\partial V/\partial R|_{R=R_m}=0$ and  $\partial^2 V/\partial R^2|_{R=R_m}=G\mu M/R_m^3>0$,
we find the relative equilibrium distance between two stars corresponding to the minimum of $V$:
\begin{eqnarray}
R_m=\frac{L^2}{G\mu^2 M}.
\label{eq_pot4}
\end{eqnarray}
Using Eq. (\ref{eq_pot4}) and $$P_{orb}=2\pi[R_{m}^3/(GM)]^{1/2}$$ for the period of orbital rotation of the   di-star system
(Kepler's third law),
we obtain
$$L=L_i=\frac{2\pi\mu R_{m}^2}{P_{orb}}=\mu(GMR_m)^{1/2}$$ and
derive the simple expression
\begin{eqnarray}
V(R_m)=-\frac{GM_1M_2}{2R_m}=-\beta G(1-\eta^2)^3/2,
\label{eq_pot5}
\end{eqnarray}
where
$$\beta=\frac{GM^5}{64L_i^2}=\frac{GM_{\odot}^5}{2L_i^2}\left(\frac{M}{2M_{\odot}}\right)^{5}$$ coincides with $\beta$ from Sect. 2
because
$$L_i^2=\mu_i^2GMR_{m,i}=\mu_i^2\left(\frac{G^{4}M^{4}P_{orb,i}^2}{4\pi^2}\right)^{1/3}.$$
Here, the relation $L=L_i$ follows from Eqs. (9) and (\ref{eq_pot4}).


\begin{figure} [ht]
\centering
{\includegraphics[width=0.49\linewidth]{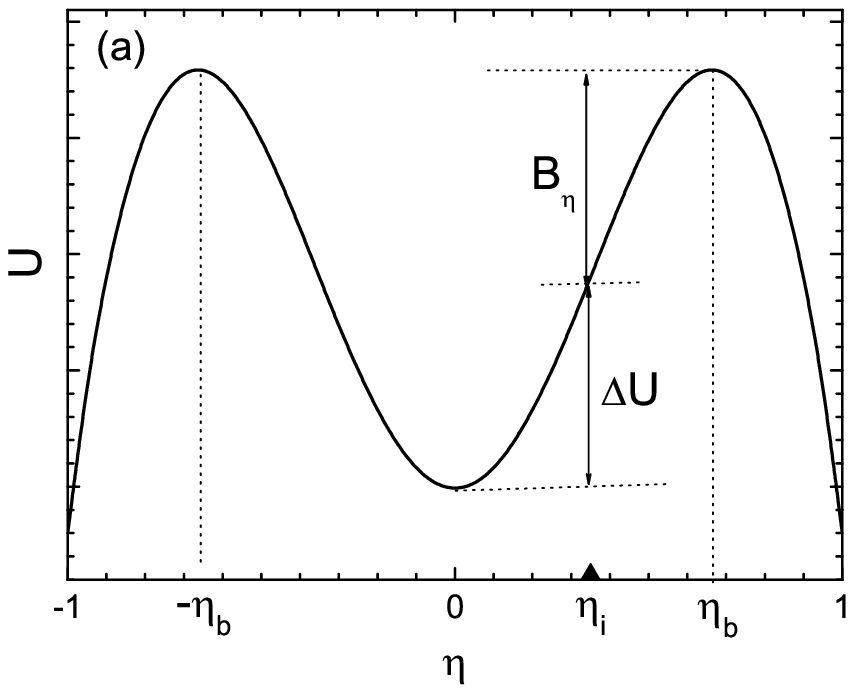}
\includegraphics[width=0.49\linewidth]{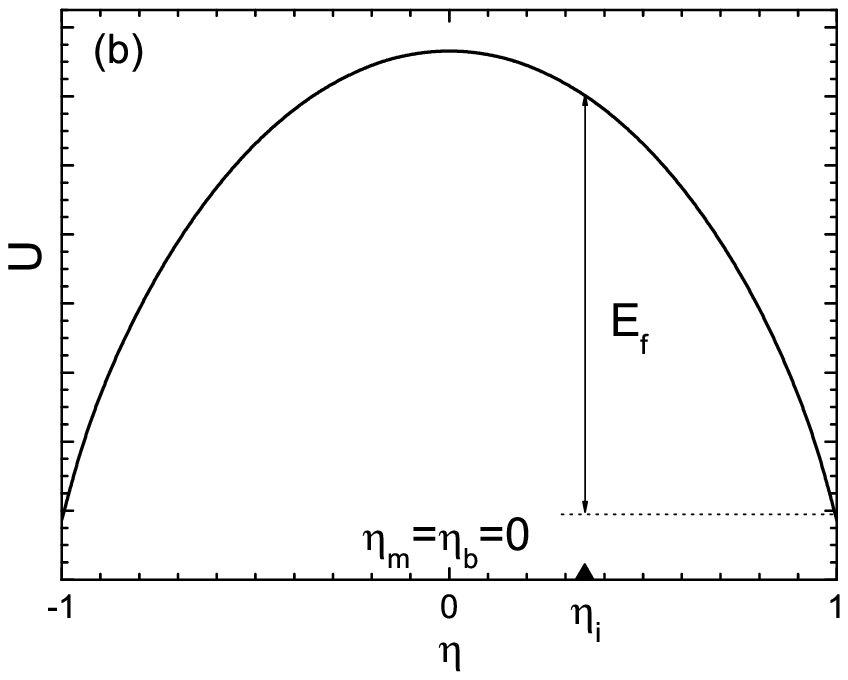}}
\caption{The schematical drawings of
the driving potential energy
of the star-star system at  $\alpha<\alpha_{cr}$ (a), and $\alpha>\alpha_{cr}$ (b).
The arrows on $x$-axis show the corresponding
initial binary stars.
The notations used in the text are indicated.
}
\label{1_fig}
\end{figure}

\begin{figure}[ht]
   \resizebox{\hsize}{!}
{\includegraphics[width=0.5\linewidth]{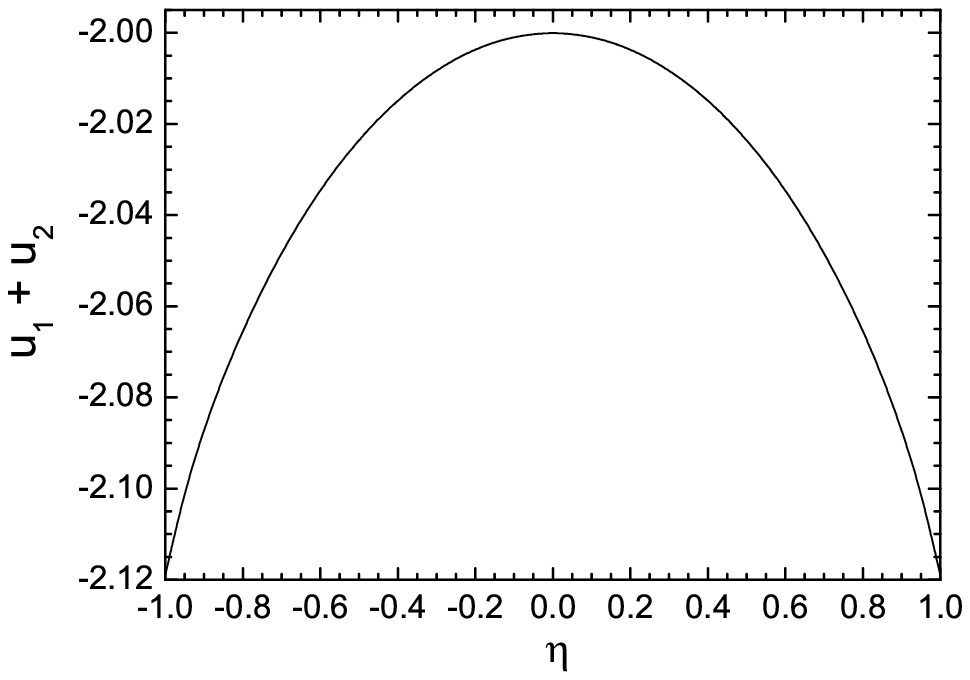}
\includegraphics[width=0.49\linewidth]{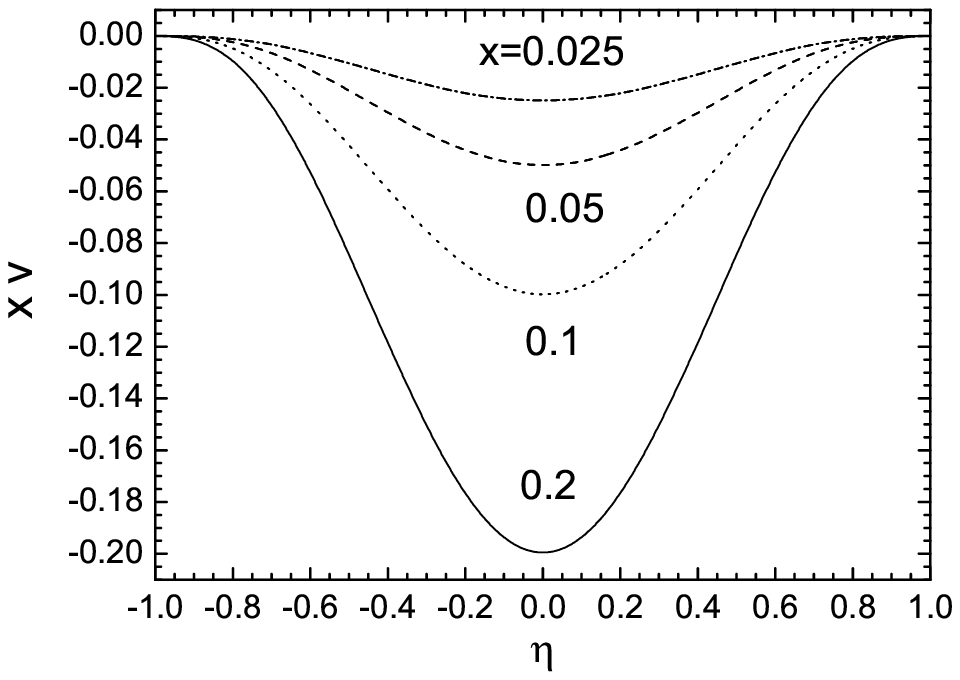}}
{\includegraphics[width=0.5\linewidth]{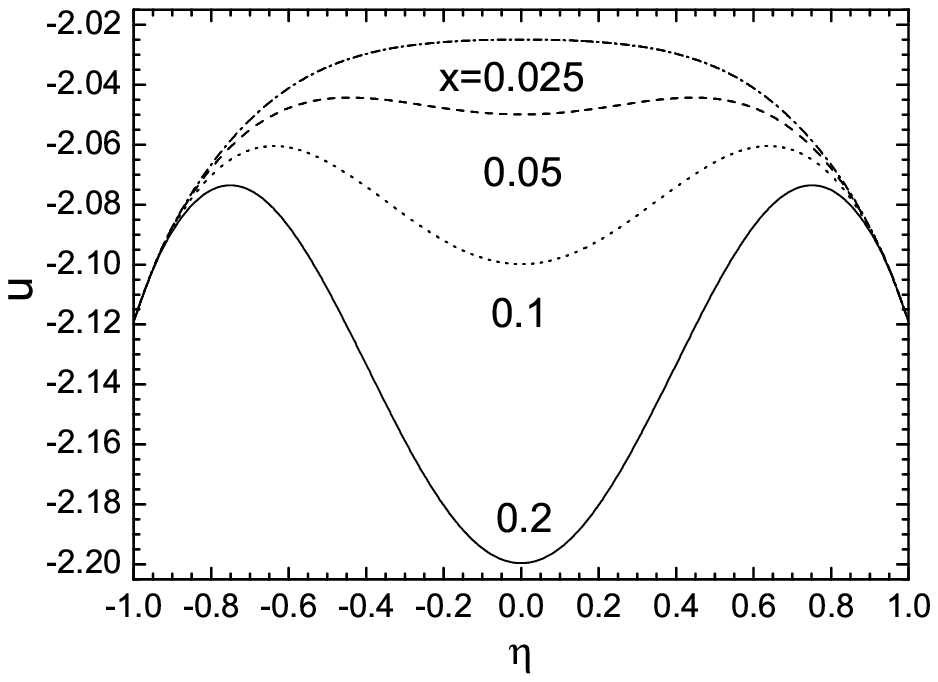}}
\caption{The calculated dimensionless  $u$, $u_1+u_2$, and $xv$
vs $\eta$ at indicated $x$.
}
\label{2_fig}
\end{figure}

\begin{figure}[ht]
   \resizebox{\hsize}{!}
{\includegraphics[width=0.5\linewidth]{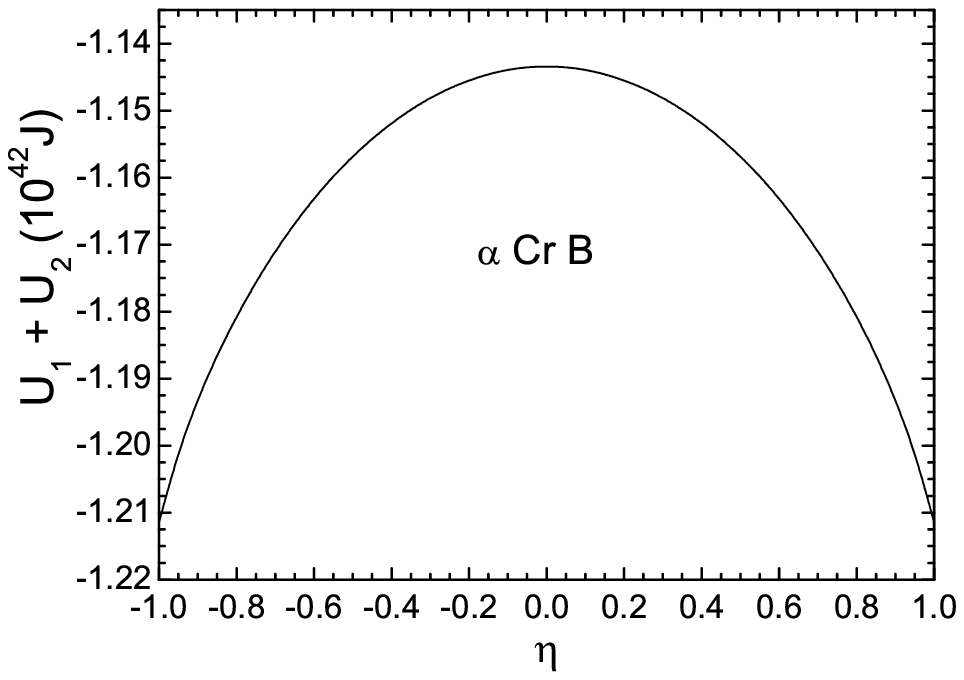}
\includegraphics[width=0.49\linewidth]{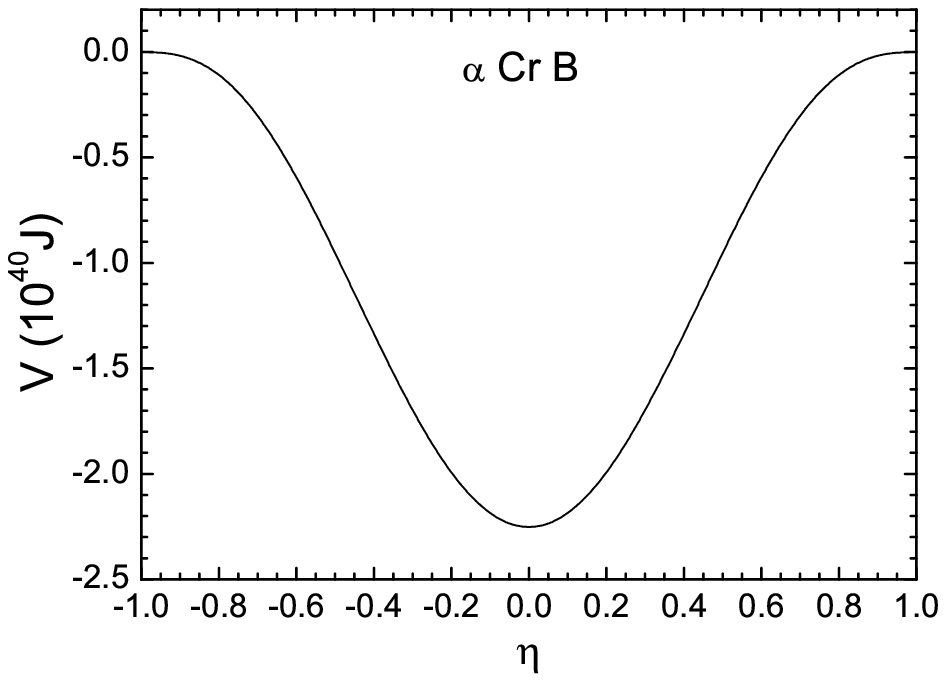}}
{\includegraphics[width=0.5\linewidth]{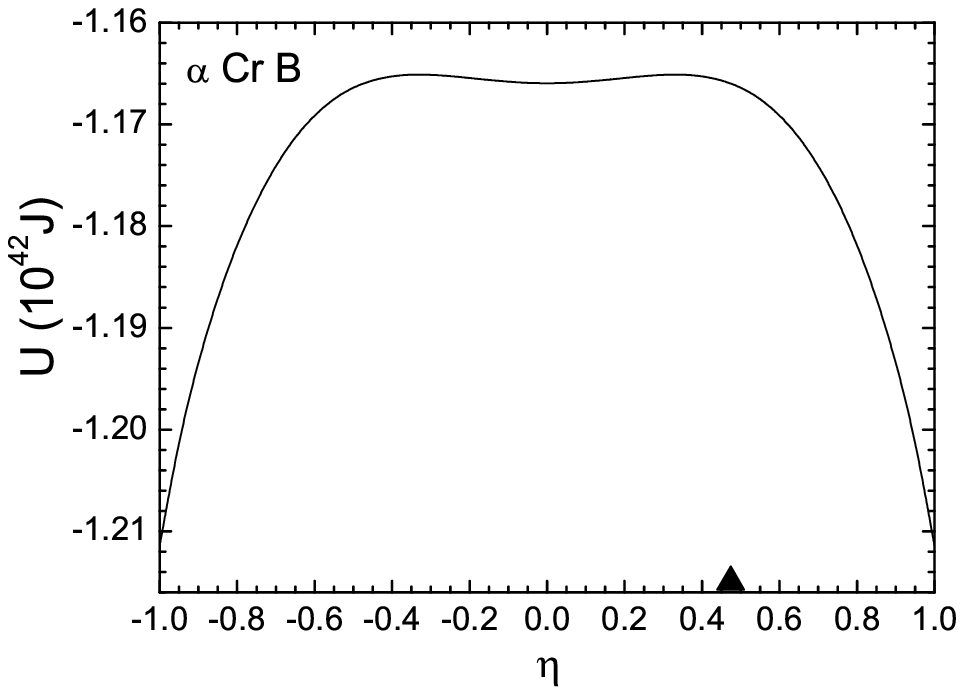}}
\caption{The calculated gravitational energy $U_1+U_2$, star-star interaction energy, and total potential energy $U$
vs $\eta$ for   close binary star $\alpha$ Cr B. The arrows on $x$-axis show the
initial $\eta_i$.
}
\label{3_fig}
\end{figure}


\begin{figure}[ht]
   \resizebox{\hsize}{!}
{\includegraphics[width=0.45\linewidth]{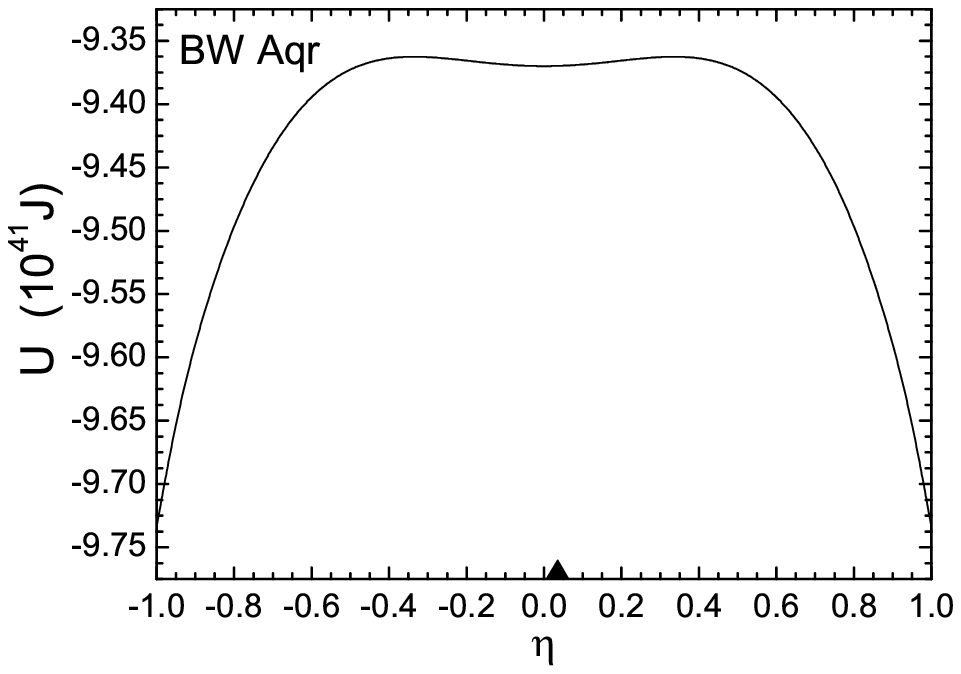}
\includegraphics[width=0.45\linewidth]{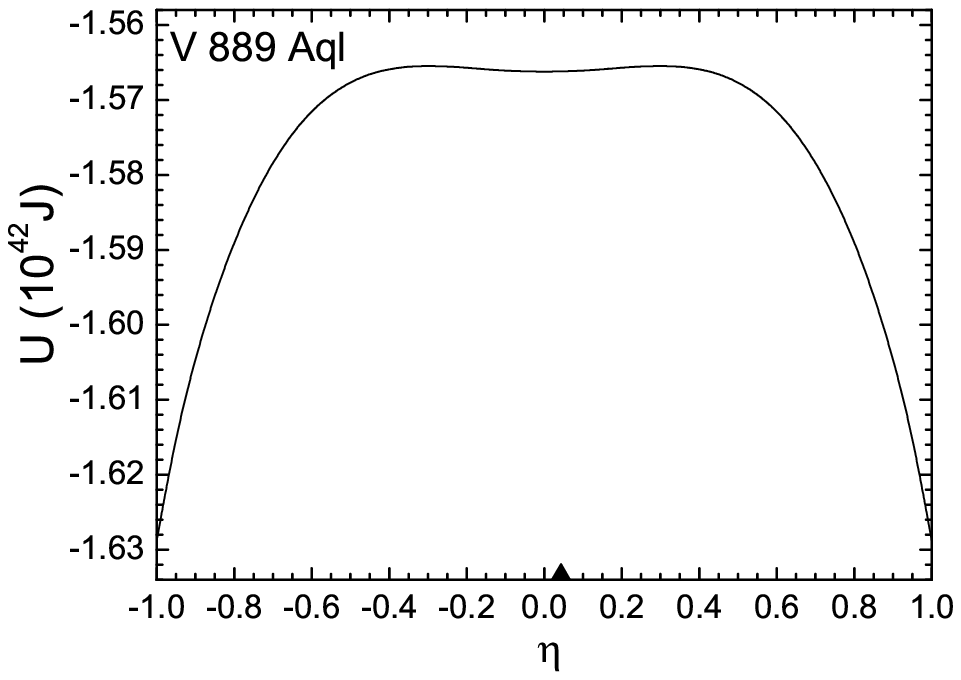}}
{\includegraphics[width=0.5\linewidth]{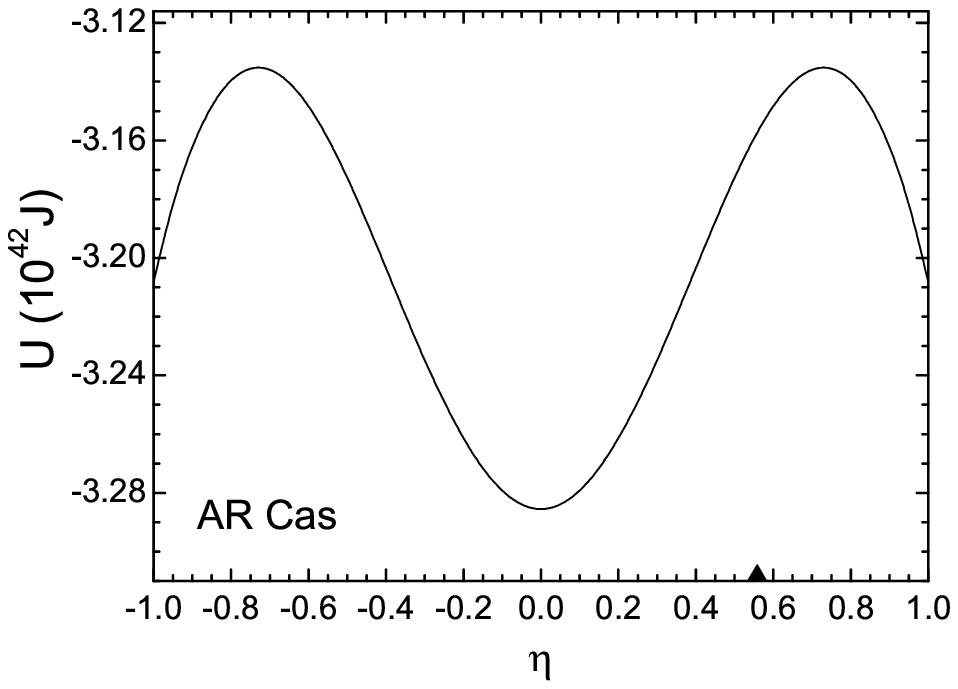}
\includegraphics[width=0.49\linewidth]{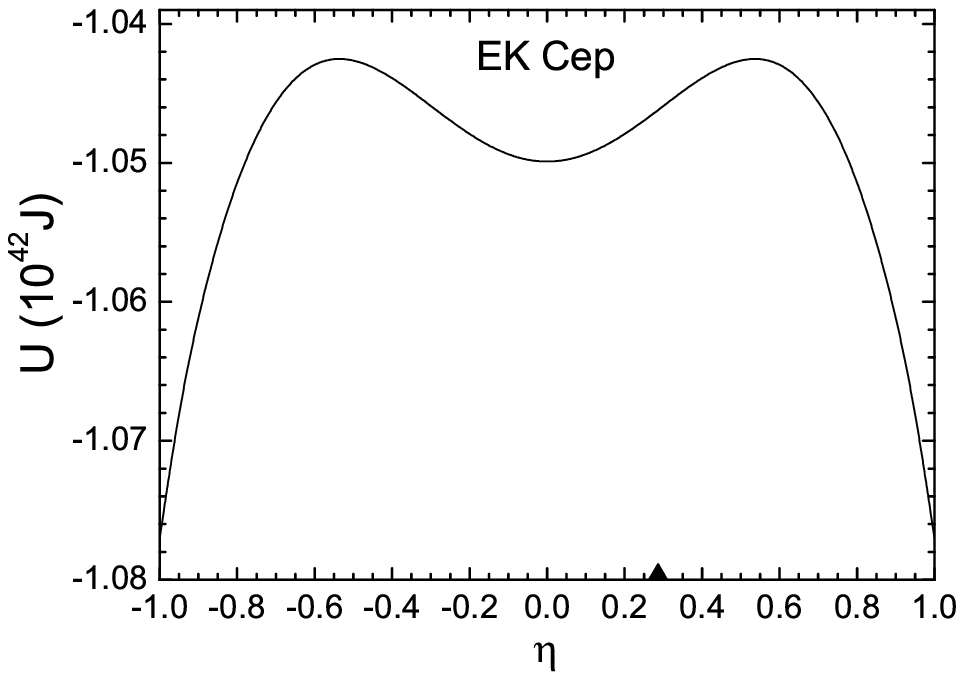}}
{\includegraphics[width=0.5\linewidth]{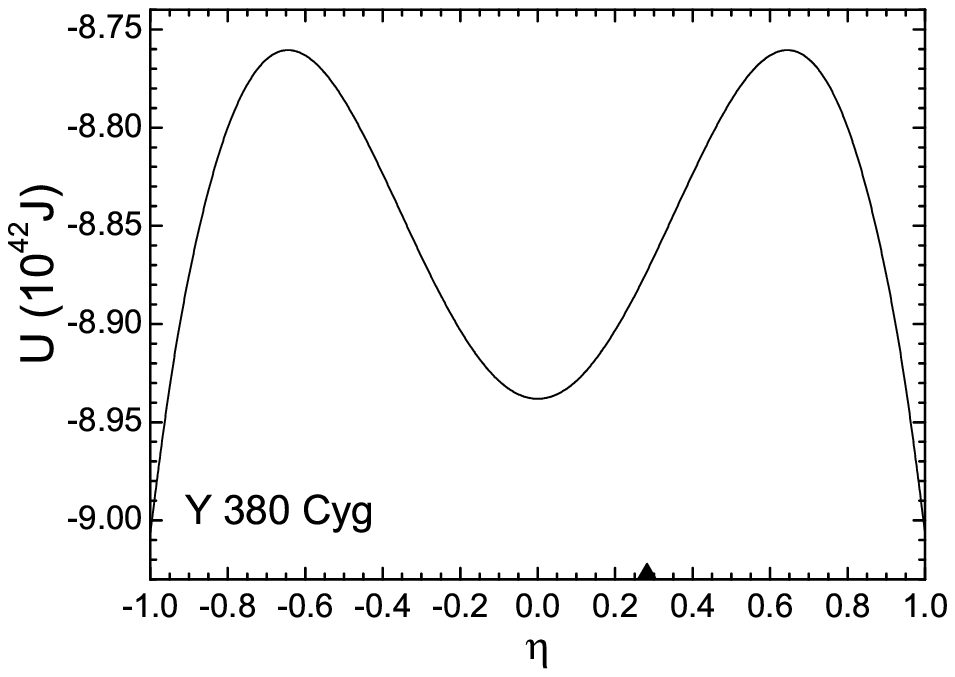}
\includegraphics[width=0.49\linewidth]{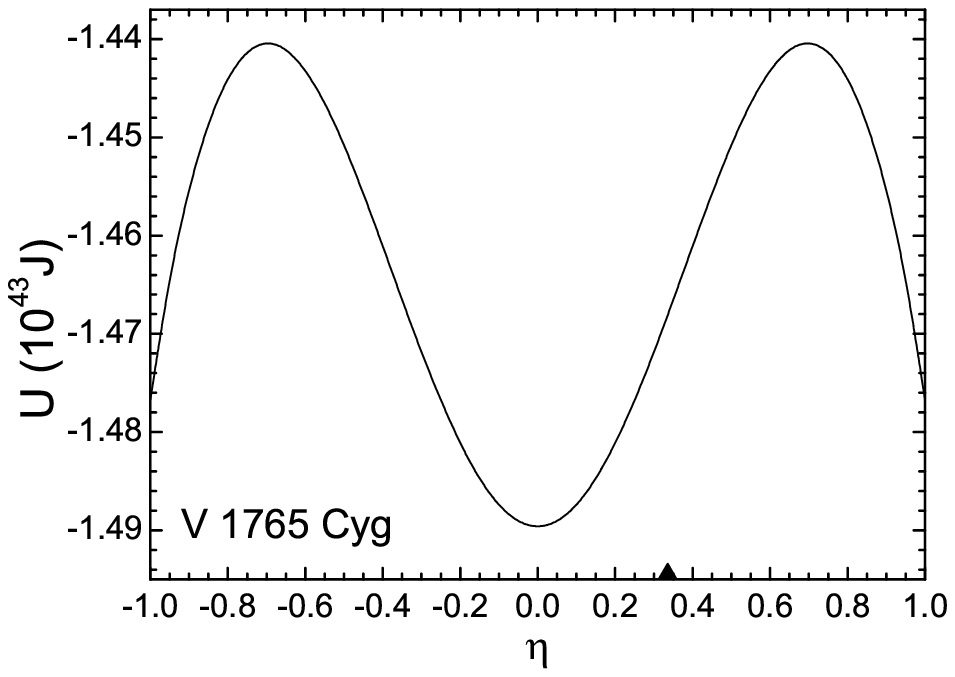}}
{\includegraphics[width=0.5\linewidth]{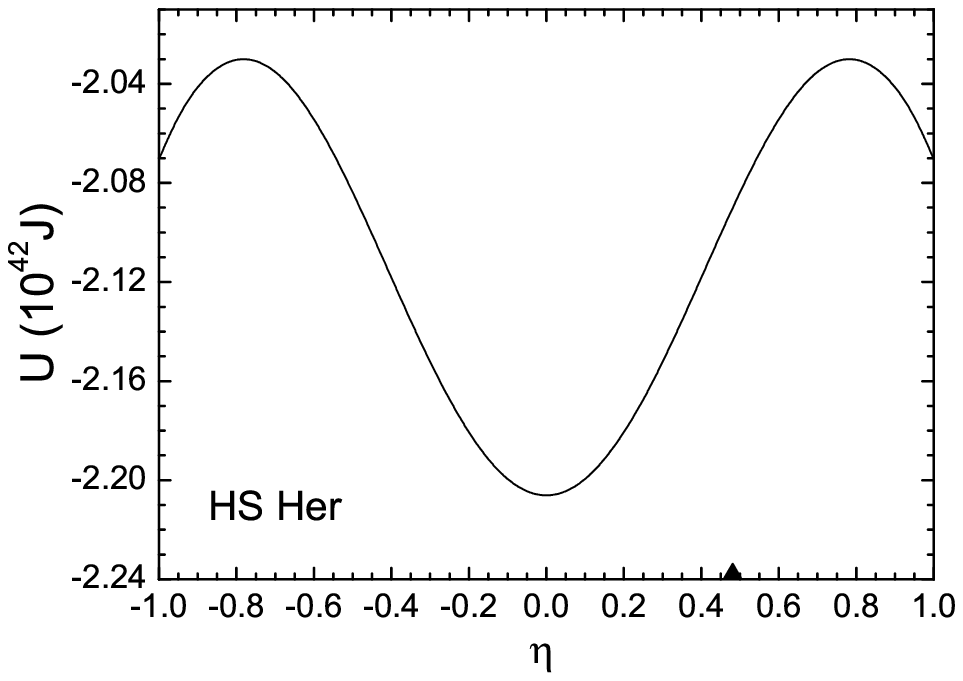}
\includegraphics[width=0.49\linewidth]{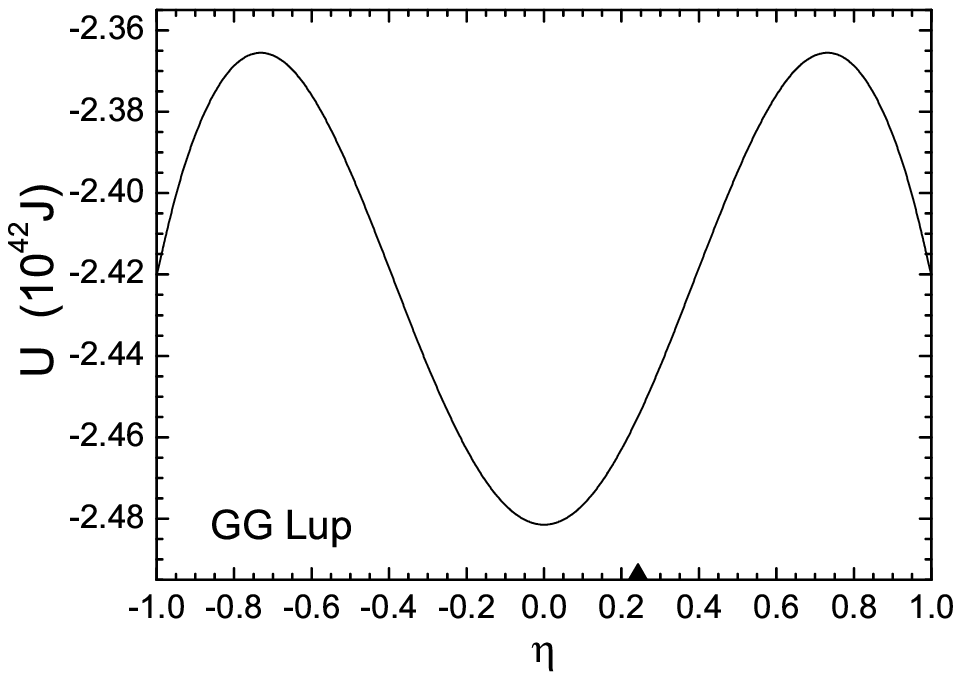}}
\caption{The calculated total potential energies $U$
vs $\eta$ for the indicated close binary stars. The arrows on $x$-axis show the corresponding
initial $\eta_i$ for binary stars.
}
\label{4_fig}
\end{figure}

\begin{figure}[ht]
   \resizebox{\hsize}{!}
{\includegraphics[width=0.45\linewidth]{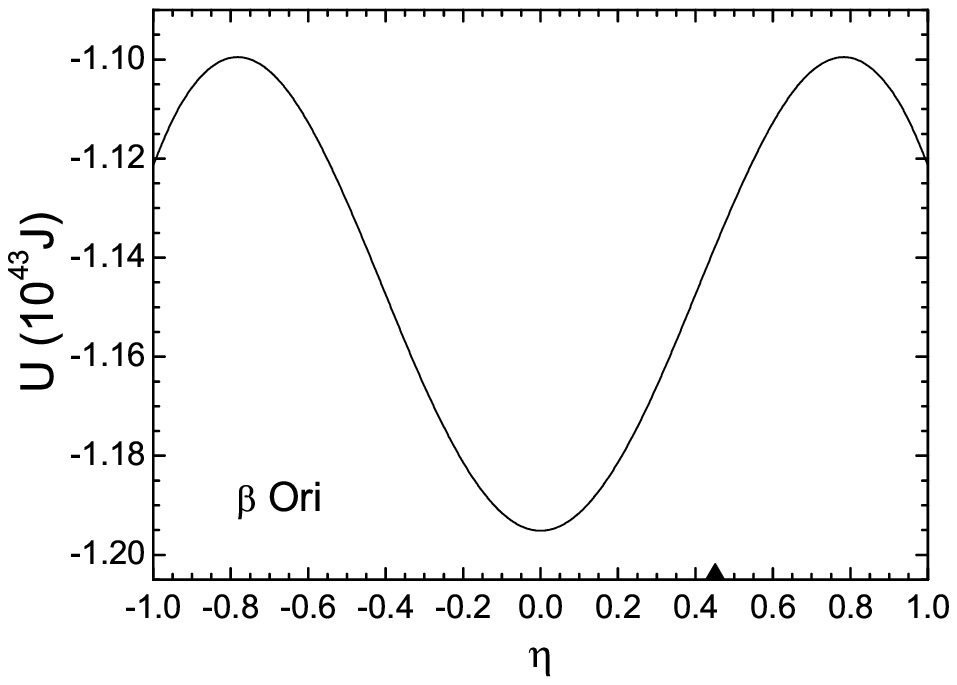}
\includegraphics[width=0.45\linewidth]{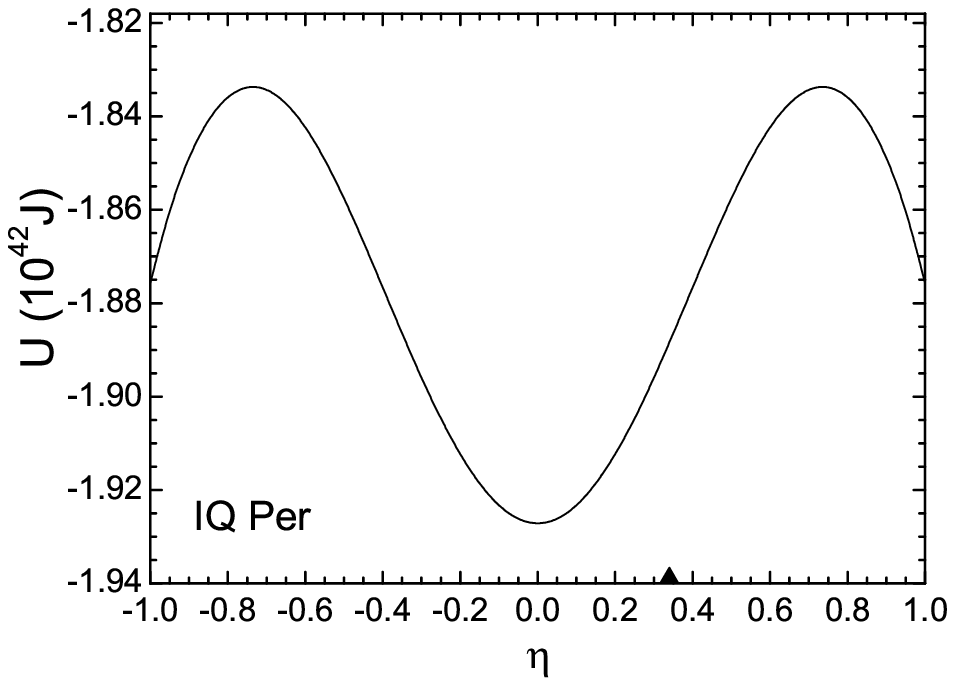}}
{\includegraphics[width=0.5\linewidth]{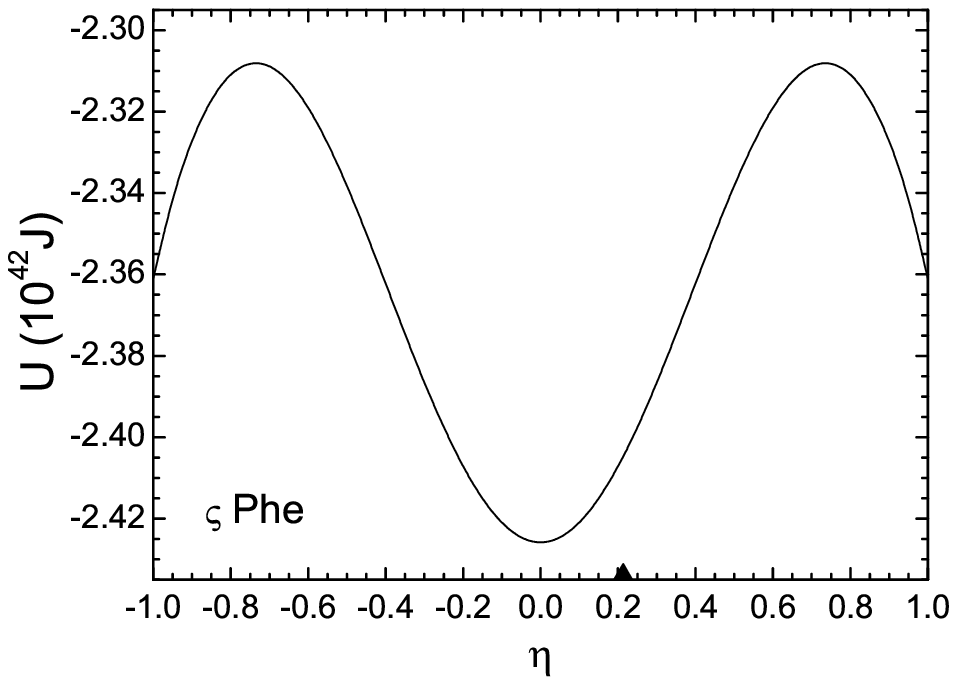}
\includegraphics[width=0.49\linewidth]{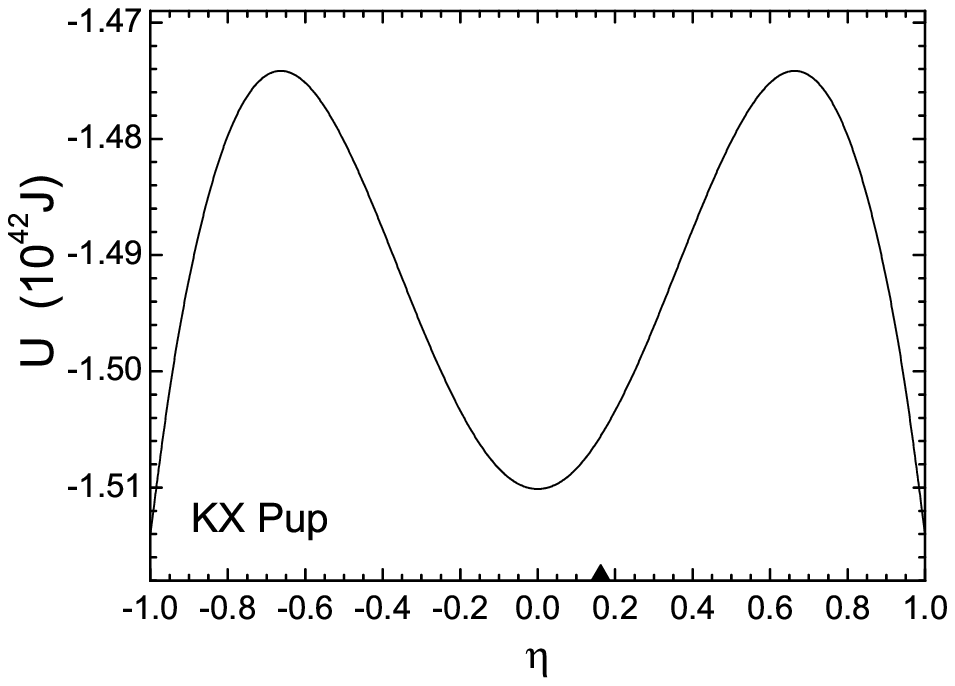}}
{\includegraphics[width=0.5\linewidth]{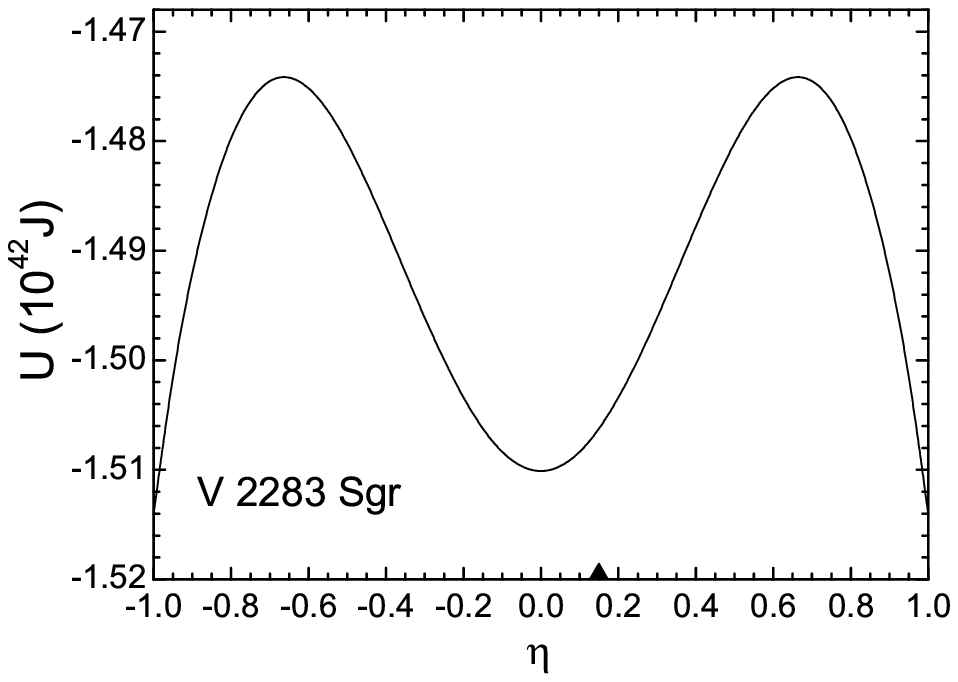}
\includegraphics[width=0.49\linewidth]{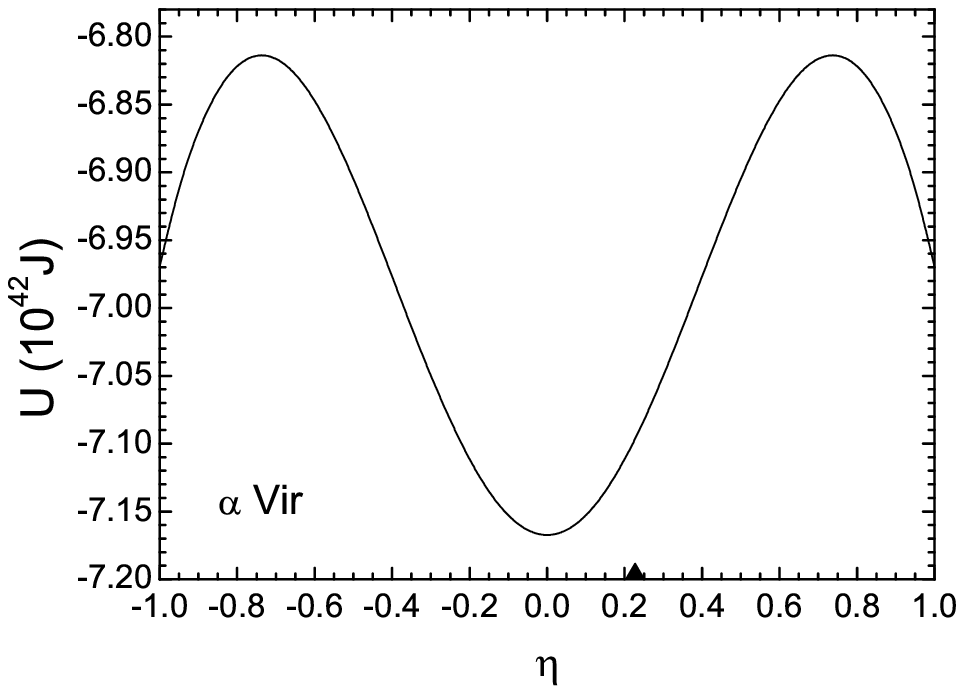}}
{\includegraphics[width=0.5\linewidth]{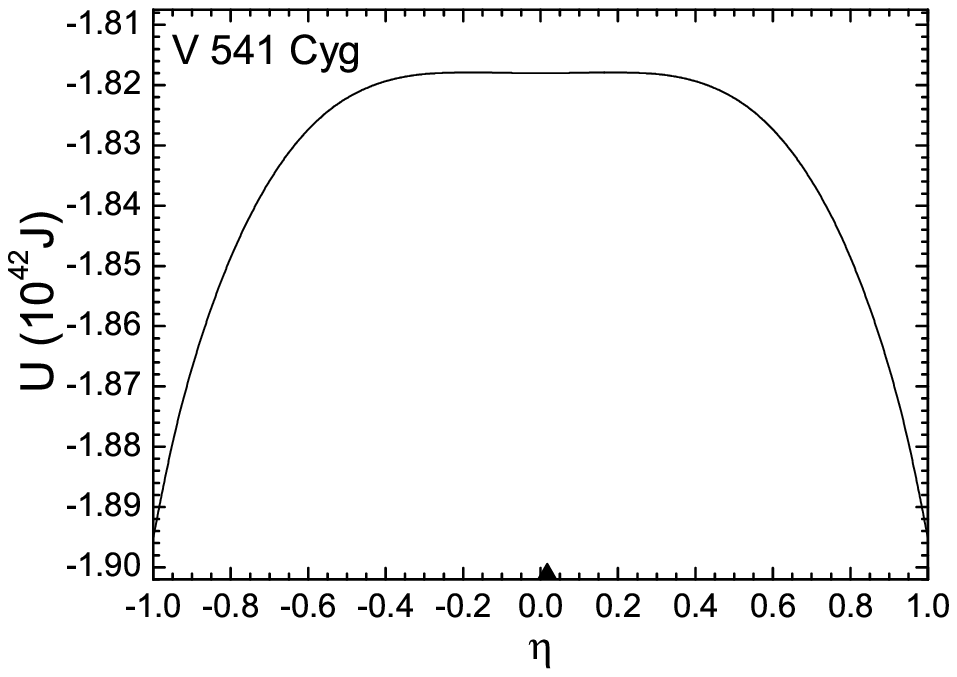}
\includegraphics[width=0.49\linewidth]{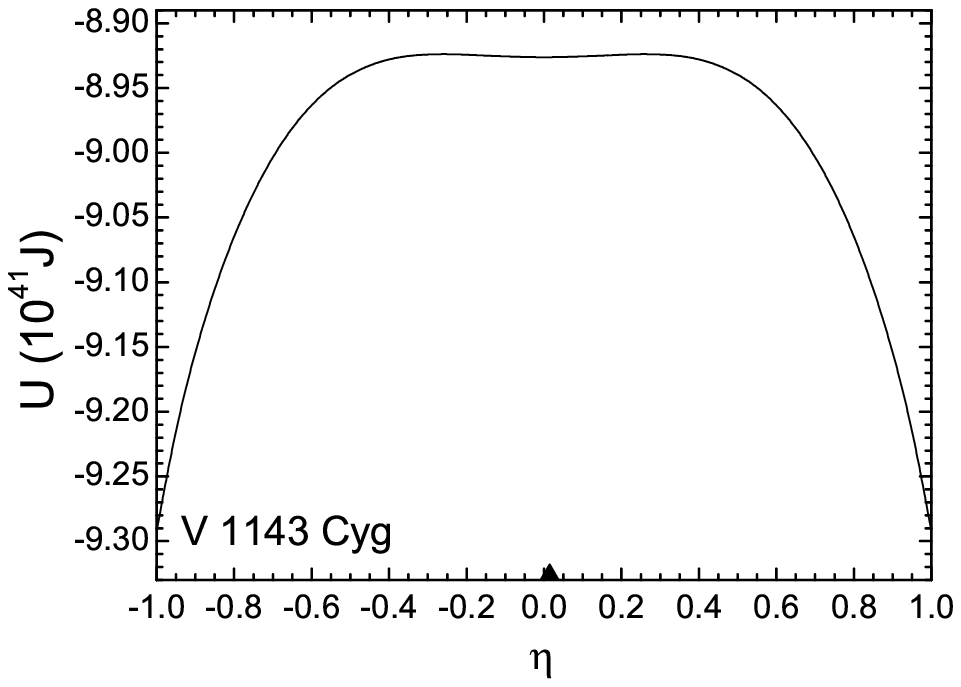}}
\caption{The same as in Fig. \ref{4_fig}, but
 for other indicated close binary stars.
}
\label{5_fig}
\end{figure}

\begin{figure}[ht]
   \resizebox{\hsize}{!}
{\includegraphics[width=0.5\linewidth]{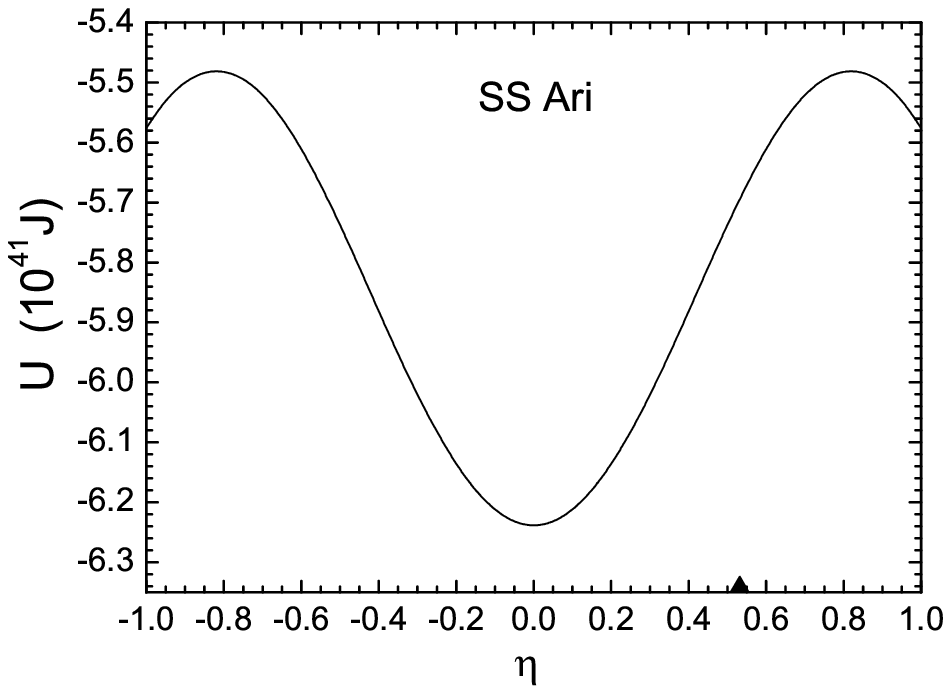}
\includegraphics[width=0.49\linewidth]{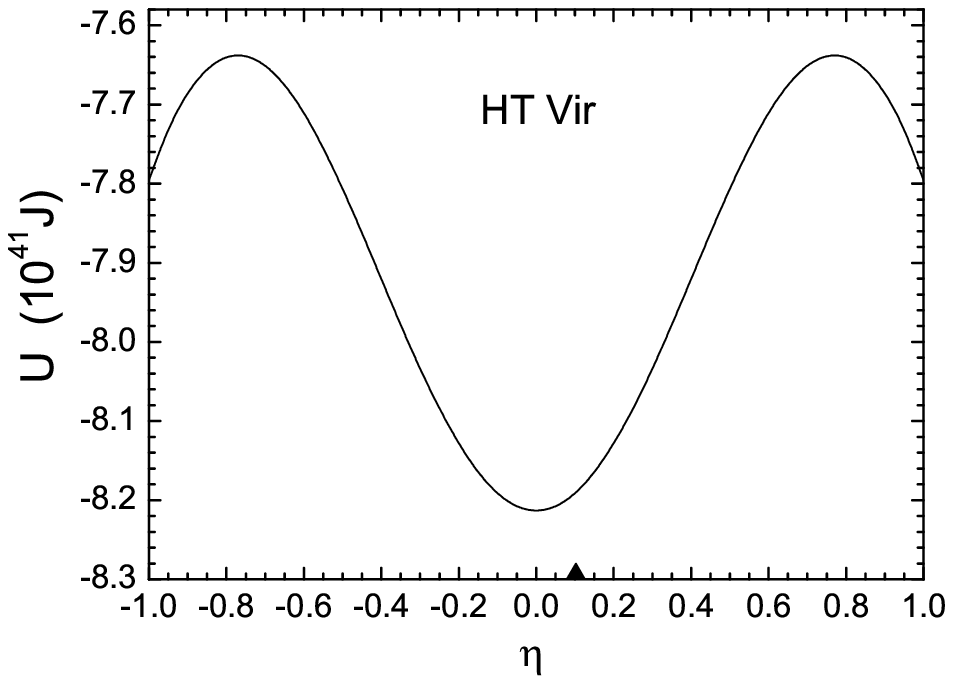}}
{\includegraphics[width=0.5\linewidth]{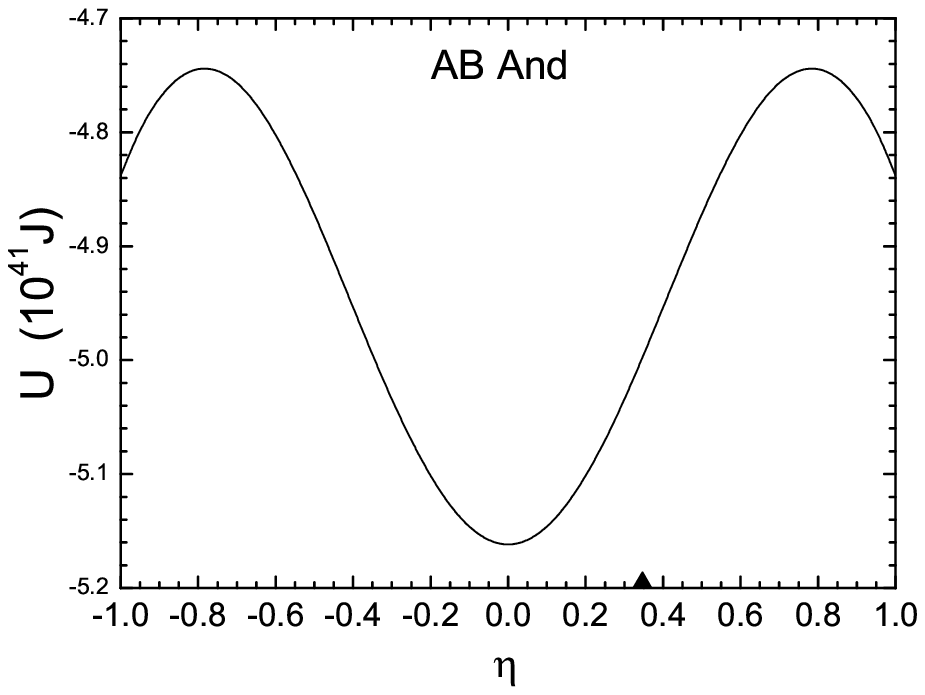}
\includegraphics[width=0.49\linewidth]{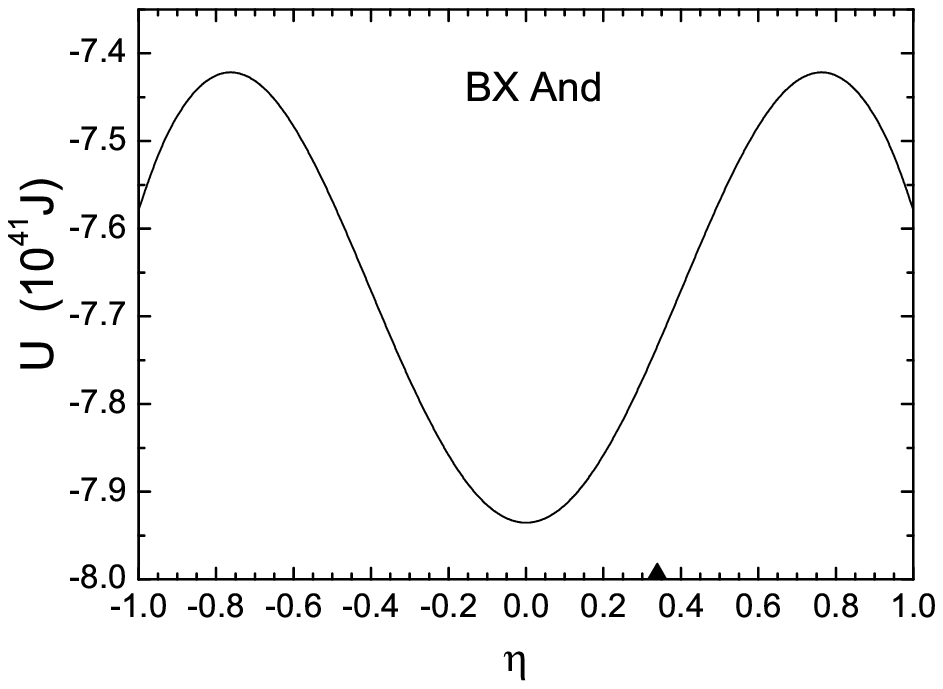}}
{\includegraphics[width=0.5\linewidth]{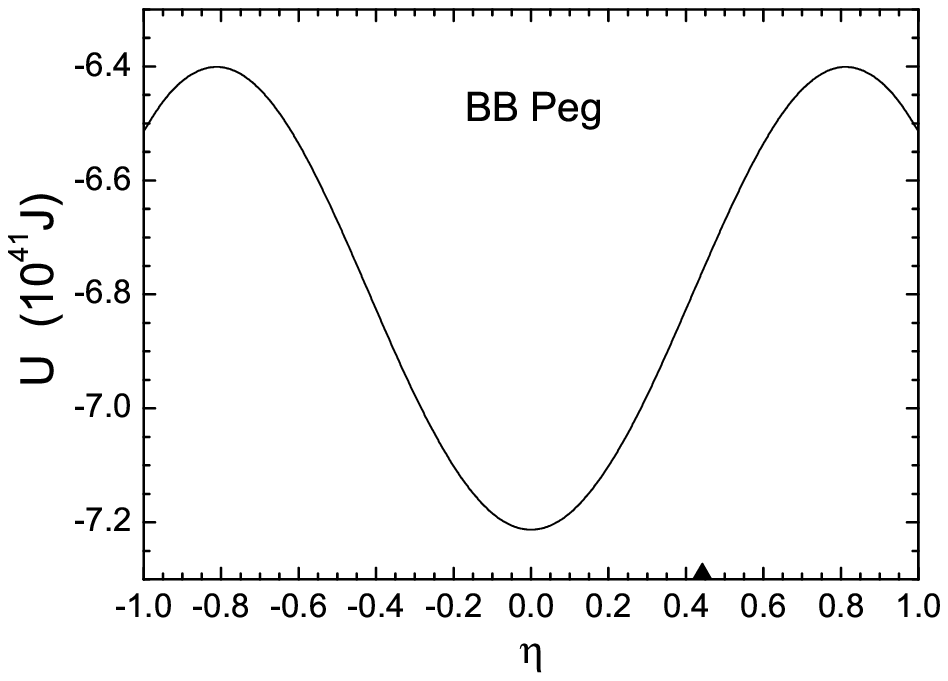}
\includegraphics[width=0.49\linewidth]{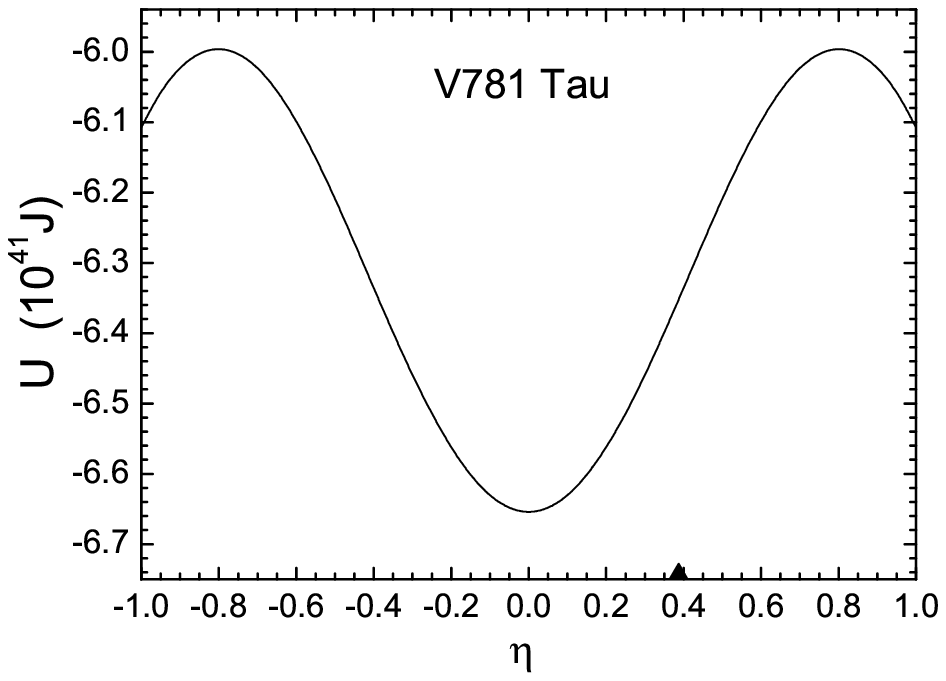}}
{\includegraphics[width=0.5\linewidth]{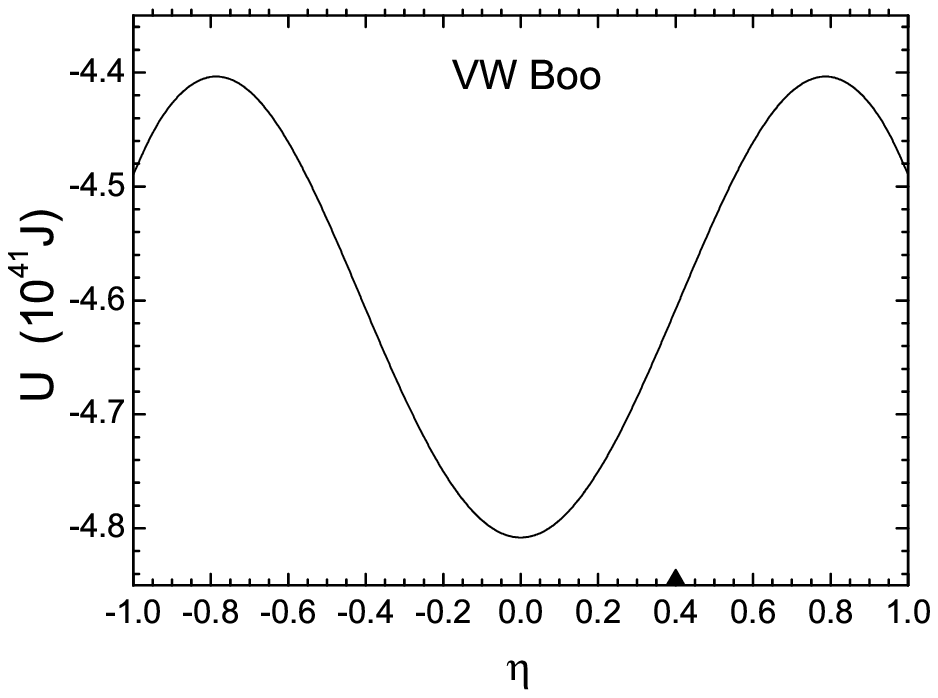}
\includegraphics[width=0.49\linewidth]{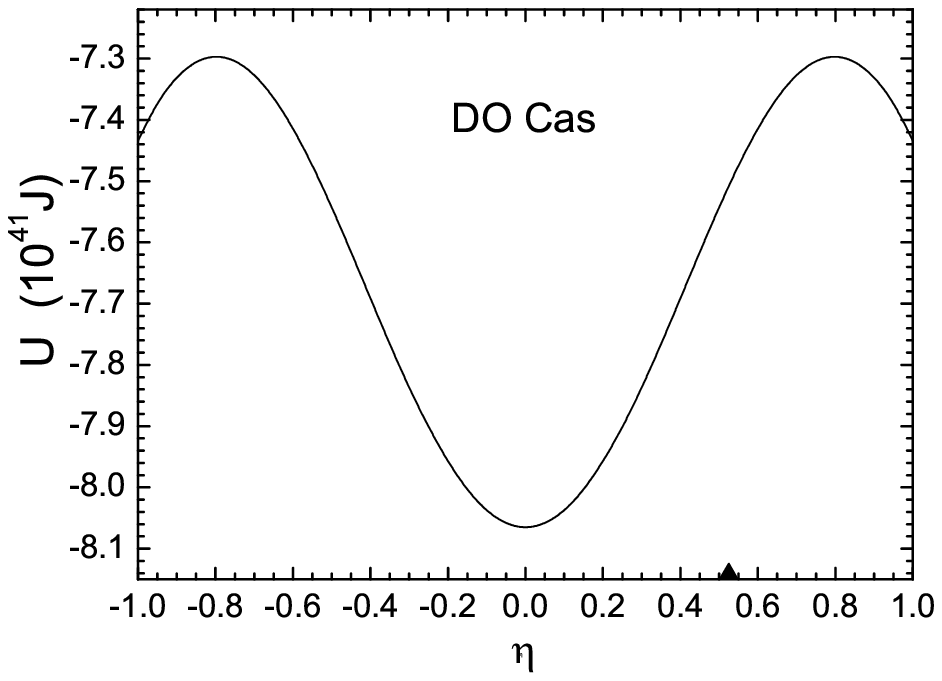}}
\caption{The same as in Fig. \ref{4_fig}, but
 for other indicated close binary stars.
}
\label{6_fig}
\end{figure}

\begin{figure}[ht]
   \resizebox{\hsize}{!}
{\includegraphics[width=0.48\linewidth]{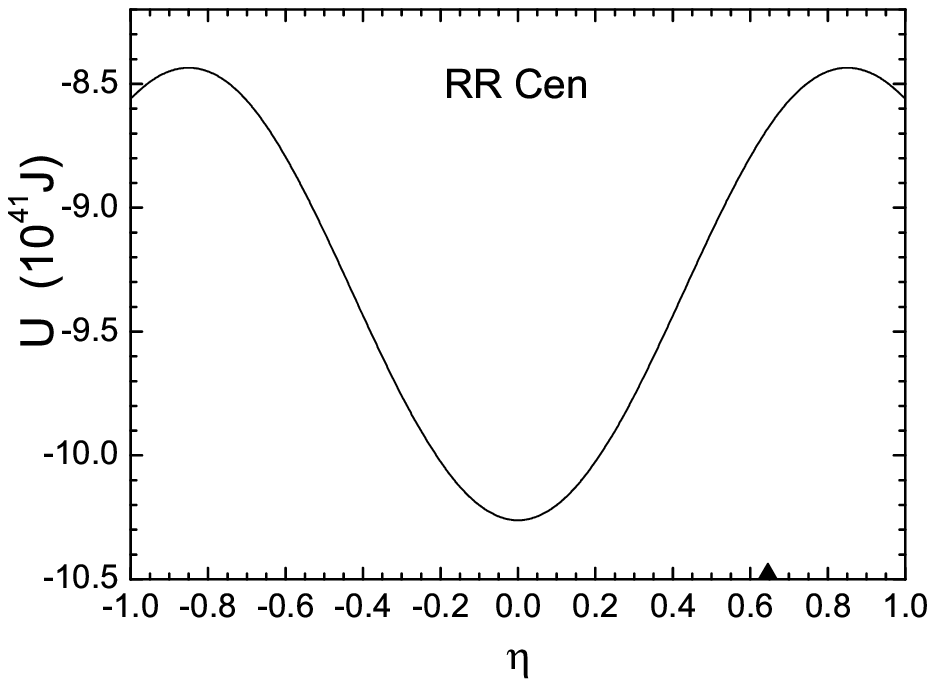}
\includegraphics[width=0.49\linewidth]{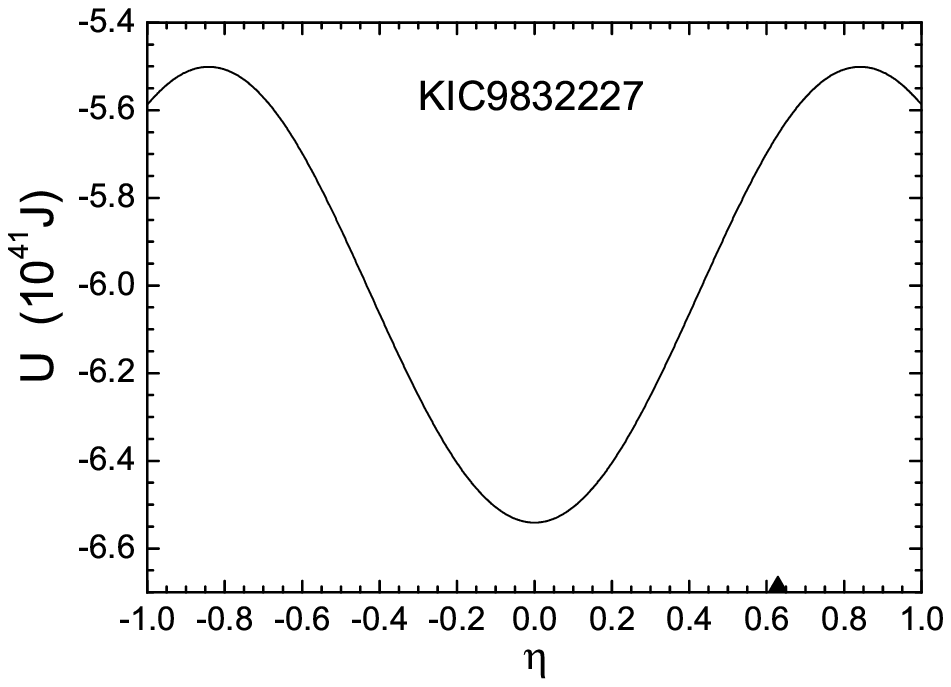}}
{\includegraphics[width=0.48\linewidth]{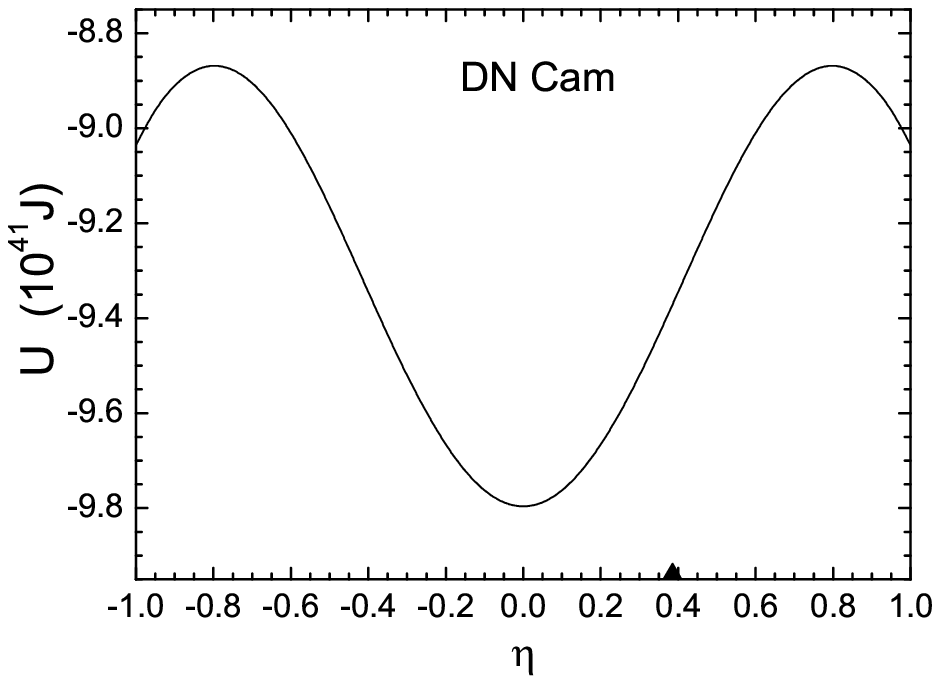}
\includegraphics[width=0.49\linewidth]{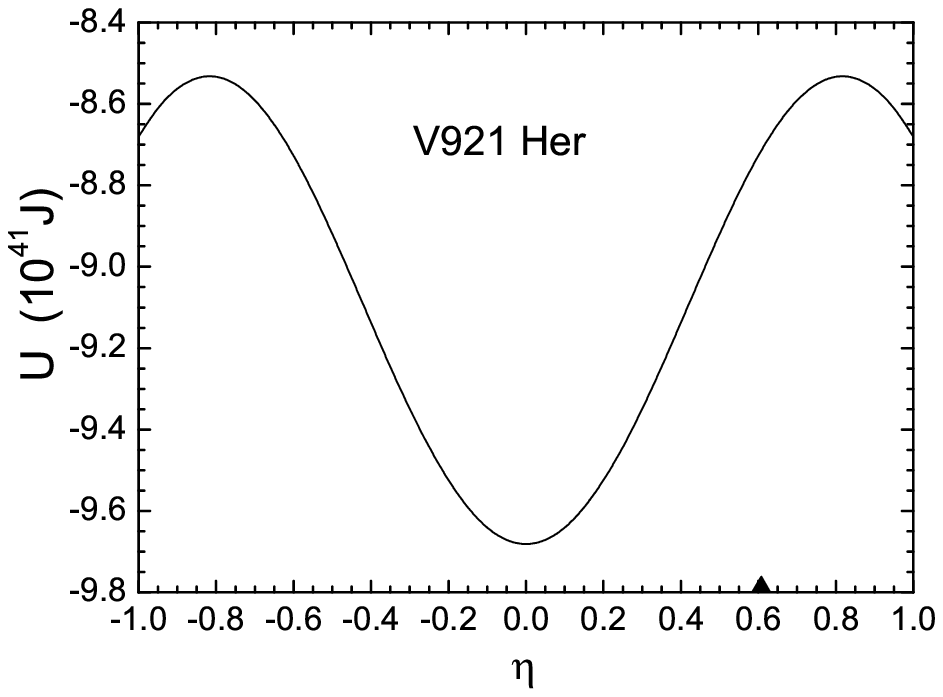}}
{\includegraphics[width=0.48\linewidth]{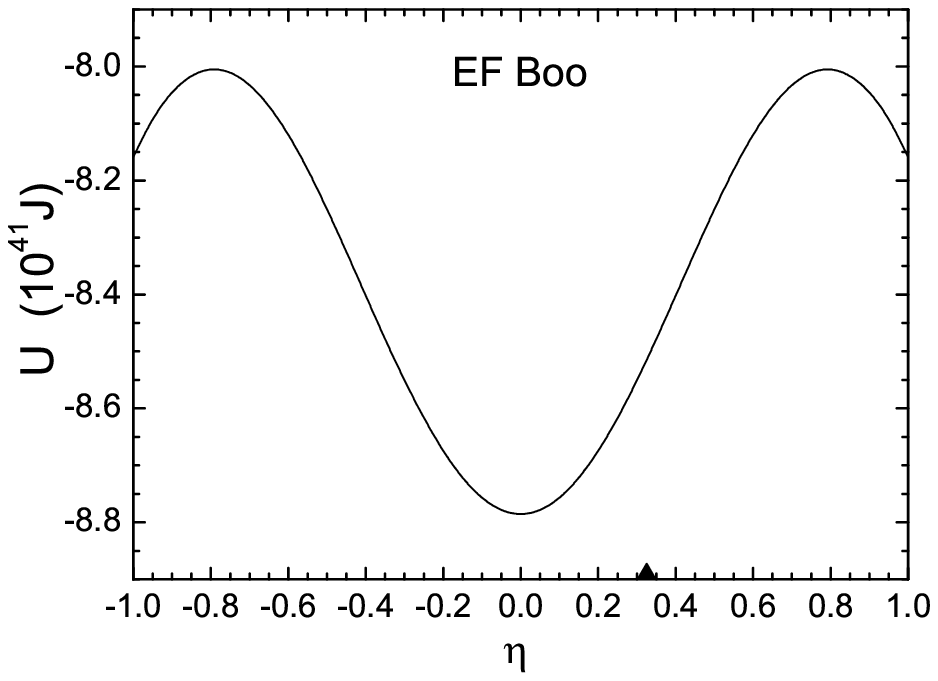}
\includegraphics[width=0.49\linewidth]{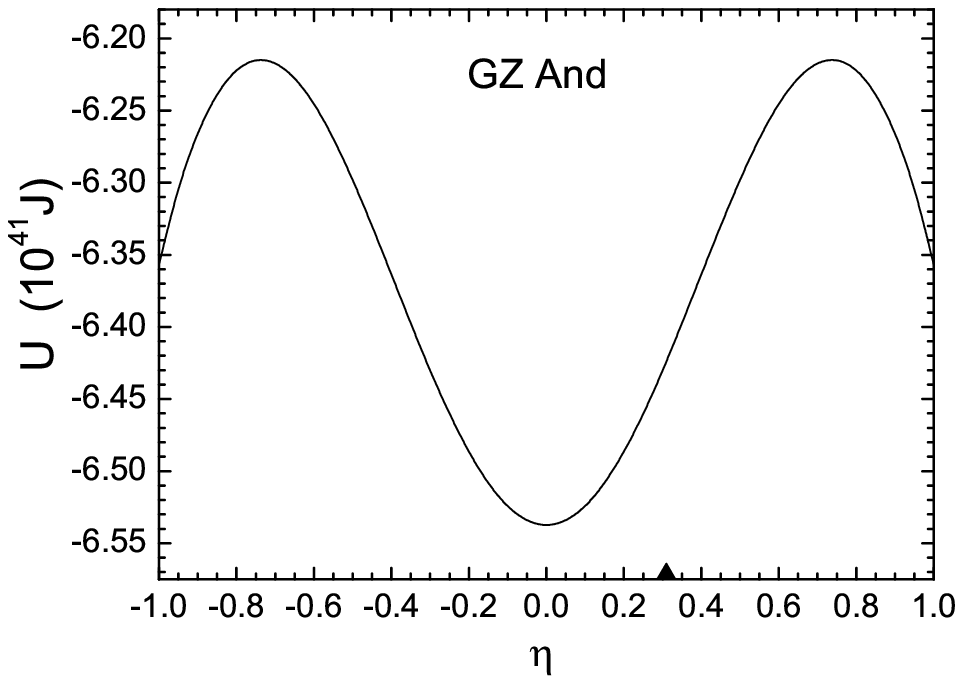}}
{\includegraphics[width=0.49\linewidth]{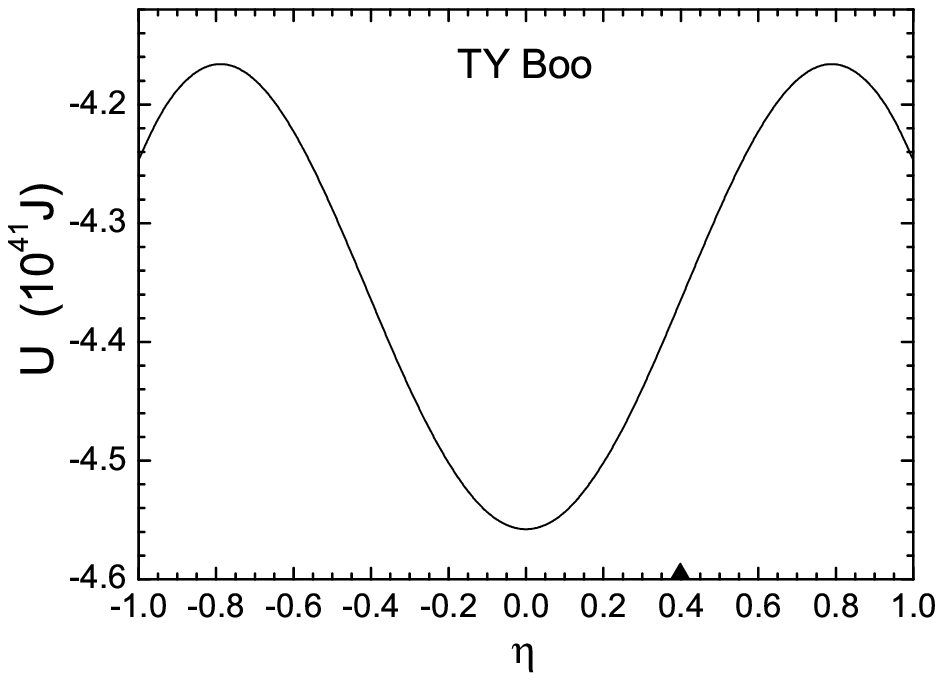}
\includegraphics[width=0.49\linewidth]{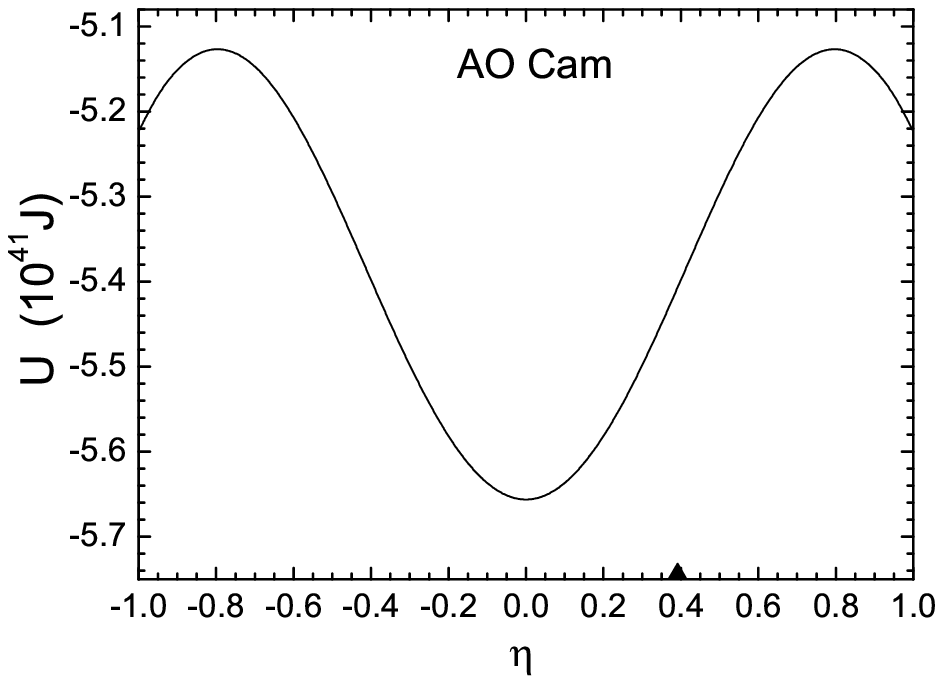}}
\caption{The same as in Fig. \ref{4_fig}, but
 for other indicated close binary stars.
}
\label{7_fig}
\end{figure}




\newpage

\begin{table}
\small\addtolength{\tabcolsep}{0pt}
\caption{The calculated  $\omega_1$, $\omega_2$,  $\beta/\alpha$,    $\Delta U=U(\eta_i)-U(\eta=0)$,
 $B_\eta=U(\eta_b)-U(\eta_i)$, and observed data $M_1/M_{\odot}$,  $M_2/M_{\odot}$ \protect\cite{Khal:2004,Vasil:2012}
for the close binary stars indicated.
}
\begin{tabular*}{\textwidth}{@{\extracolsep{\fill}}|c|c|c|c|c|c|c|c|}
\hline
 Di-star & $\frac{M_1}{M_{\odot}}$ & $\frac{M_2}{M_{\odot}}$ & $\omega_1$  &  $\omega_2$& $\beta/\alpha$ &  $\Delta U$    (J)    &   $B_\eta$   (J)    \\
\hline
BW Aqr   & 1.48   &  1.38 &   1.49 & 1.5    & 0.040  & $2\times 10^{37}$      &   $7\times 10^{38}$                \\
V 889 Aql& 2.4    &  2.2  &   1.32 & 1.35   & 0.038  &   $3\times 10^{37}$      &   $8\times 10^{38}$                  \\
V 539 Ara& 6.24   &  5.31 &   1.04 & 1.08   &  0.150  &  $5\times 10^{39}$      &   $2\times 10^{41}$                   \\
AS Cam   & 3.31   &  2.51 &   1.22 & 1.31   & 0.098  &   $4\times 10^{39}$      &   $3\times 10^{40}$                 \\
EM  Car  & 22.8   & 21.4  & 0.75   & 0.76   & 0.309 &   $8\times 10^{39}$      &   $2\times 10^{42}$           \\
GL Car   & 13.5   & 13.0  & 0.86   & 0.87   & 0.287  &   $10^{39}$              &   $10^{42}$                      \\
QX Car   & 9.27   & 8.48  & 0.94   & 0.96   &  0.152  &  $2\times 10^{39}$      &   $3\times 10^{41}$               \\
AR Cas   & 6.7    & 1.9   & 1.02   & 1.40   &  0.170  &  $10^{41}$              &   $2\times 10^{40}$           \\
IT Cas   & 1.40   & 1.40  & 1.51   & 1.51   &  0.056  &        0                &   $4\times 10^{39}$              \\
OX Cas   & 7.20   & 6.30  & 1      & 1.04   &  0.192 & $5\times 10^{39}$       &   $3\times 10^{41}$            \\
PV Cas   & 2.79   & 2.79  & 1.27   & 1.27   &  0.144  &          0              &   $7\times 10^{40}$           \\
KT Cen   & 5.30   & 5.00  & 1.08   & 1.27   &  0.116  & $4\times 10^{38}$       &   $10^{41}$                     \\
V 346 Cen& 11.8   & 8.4   & 0.89   & 0.97   &  0.137  & $3\times 10^{40}$       &   $2\times 10^{41}$            \\
CW Cep   & 11.60  & 11.10 & 0.89   & 0.9    &  0.243  & $10^{39}$               &   $7\times 10^{41}$            \\
EK Cep   & 2.02   & 1.12  &  1.38  &  1.6   &  0.066  & $4\times 10^{39}$       &   $4\times 10^{39}$             \\
Y Cyg    & 17.5   & 17.3  & 0.8    & 0.81   &  0.292  & $2\times 10^{38}$       &   $2\times 10^{41}$                \\
Y 380 Cyg& 14.3   & 8.0   & 0.85   & 1.27   &  0.103 & $7\times 10^{40}$       &   $10^{41}$                    \\
V 453 Cyg& 14.5   & 11.3  & 0.84   &  0.9   &  0.213  & $4\times 10^{40}$       &   $7\times 10^{41}$               \\
V 477 Cyg& 1.79   & 1.35  & 1.42   &  1.53  &  0.088  & $2\times 10^{39}$       &   $10^{40}$                       \\
V 478 Cyg& 16.6   & 16.3  &  0.81  &  0.82  &  0.290  & $4\times 10^{38}$       &   $10^{42}$                       \\
V 541 Cyg&  2.69  & 2.60  &  1.28  &  1.29  &  0.033  & $2\times 10^{36}$       &   $10^{38}$                        \\
V 1143 Cyg& 1.39  & 1.35  &  1.51  &  1.53  &  0.036  & $2\times 10^{36}$       &   $2\times 10^{38}$               \\
V 1765 Cyg& 23.5  & 11.7  &  0.75  &  0.89  & 0.138  &  $2\times 10^{41}$       &   $3\times 10^{41}$              \\
CO Lac    & 3.13  & 2.75  &  1.24  &  1.28  & 0.163  &  $2\times 10^{39}$       &   $9\times 10^{40}$              \\

\hline
\end{tabular*}
\label{tab1}
\end{table}

\begin{table}
\small\addtolength{\tabcolsep}{0pt}
\caption{The same as in Table I, but for other close binary stars.
}
\begin{tabular*}{\textwidth}{@{\extracolsep{\fill}}|c|c|c|c|c|c|c|c|}
\hline
 Di-star & $\frac{M_1}{M_{\odot}}$ & $\frac{M_2}{M_{\odot}}$ & $\omega_1$  &  $\omega_2$& $\beta/\alpha$ &  $\Delta U$   (J)   &   $B_\eta$   (J)  \\
\hline
DI Her    & 5.15  & 4.52  &  1.09  &  1.13  & 0.060  &  $7\times 10^{38}$       &   $2\times 10^{40}$            \\
HS Her    & 4.25  & 1.49  &  0.89  &  1.49  & 0.258  &  $10^{41}$               &   $10^{41}$                        \\
GG Lup    & 4.12  & 2.51  &  1.15  &  1.3   & 0.173  &  $3\times 10^{40}$       &   $10^{41}$                       \\
RU Mon    & 3.60  & 3.33  & 1.19   &  1.22  &  0.101  & $4\times 10^{38}$       &   $5\times 10^{40}$              \\
GN Nor    & 2.50  & 2.50  & 1.31   &  1.31  &  0.061  &       0                 &   $10^{40}$                     \\
U Oph     & 5.02  & 4.52  & 1.1    &  1.13  &  0.203  & $2\times 10^{39}$       &   $2\times 10^{41}$             \\
V 451 Oph & 2.77  & 2.35  & 1.27   &  1.33  &  0.119  & $2\times 10^{39}$       &   $5\times 10^{40}$              \\
$\beta$ Ori&19.8  & 7.5   & 0.87   &  0.99  &  0.259  & $6\times 10^{41}$       &   $4\times 10^{41}$              \\
FT Ori & 2.50     & 2.30  & 1.31   & 1.33   &  0.089  & $3\times 10^{38}$       &   $5\times 10^{40}$            \\
AG Per & 5.36     & 4.90  & 1.08   & 1.1    &  0.187  & $2\times 10^{39}$       &   $2\times 10^{41}$              \\
IQ Per & 3.51     & 1.73  & 1.2    & 1.43   &  0.177  & $4\times 10^{40}$       &   $6\times 10^{40}$              \\
$\varsigma$ Phe   & 3.93  & 2.55   & 1.17   & 1.3 & 0.178  &  $2\times 10^{40}$ &   $10^{41}$                      \\
KX Pup   & 2.5    & 1.8   &  1.31  & 1.42   & 0.114  &  $5\times 10^{39}$       &   $3\times 10^{40}$              \\
NO Pup   & 2.88   & 1.5   &  1.26  & 1.49   &  0.192  & $3\times 10^{40}$       &   $6\times 10^{40}$               \\
VV Pyx   & 2.10   & 2.10  & 1.37   & 1.37   & 0.064  &       0                  &   $10^{40}$                     \\
YY Sgr   & 2.36   & 2.29  & 1.33   & 1.33   &  0.099  & $4\times 10^{37}$       &   $3\times 10^{40}$              \\
V 523 Sgr& 2.1    & 1.9   & 1.37   & 1.4    &  0.098  & $3\times 10^{38}$       &   $3\times 10^{40}$              \\
V 526 Sgr& 2.11   & 1.66  & 1.36   &  1.45  &  0.111  & $2\times 10^{39}$       &   $3\times 10^{40}$              \\
V 1647 Sgr& 2.19   & 1.97   & 1.35 &  1.39  &  0.080  & $3\times 10^{38}$       &   $2\times 10^{40}$              \\
V 2283 Sgr& 3.0    & 2.22   & 1.25 &  1.25  &  0.092  & $4\times 10^{39}$       &   $3\times 10^{40}$             \\
V 760 Sco& 4.98    & 4.62   & 1.1  &  1.12  &  0.199  & $10^{39}$               &   $2\times 10^{41}$               \\
AO Vel   & 3.2    & 2.9     & 1.23 &  1.26  & 0.163  &  $10^{39}$               &   $10^{41}$                      \\
EO Vel  & 3.21    & 2.77    &  1.23&  1.27  & 0.072  &  $7\times 10^{38}$       &   $2\times 10^{40}$              \\
$\alpha$ Vir& 10.8& 6.8     &  0.91&  0.99  & 0.180  &  $7\times 10^{40}$       &   $3\times 10^{41}$              \\
DR Vul    & 13.2    & 12.1  & 1.03 &  0.88  & 0.295  &  $7\times 10^{39}$       &   $10^{42}$                     \\

\hline
\end{tabular*}
\label{tab1}
\end{table}

\begin{table}
\small\addtolength{\tabcolsep}{0pt}
\caption{The same as in Table I, but for contact binary stars.
The observed data $M_1/M_{\odot}$,  $M_2/M_{\odot}$ are from Refs. \protect\cite{Molnar:2017,Egg}.
}
\begin{tabular*}{\textwidth}{@{\extracolsep{\fill}}|c|c|c|c|c|c|c|c|}
\hline
 Di-star & $\frac{M_1}{M_{\odot}}$ & $\frac{M_2}{M_{\odot}}$ & $\omega_1$  &  $\omega_2$& $\beta/\alpha$ &  $\Delta U$   (J)   &   $B_\eta$   (J)  \\
\hline
AB And    & 1.01  & 0.49  &  1.64  &  1.96  & 0.261  &  $2\times 10^{40}$       &   $3\times 10^{40}$            \\
GZ And    & 1.12  & 0.59  &  1.60  &  1.88  & 0.283  &  $2\times 10^{40}$       &   $4\times 10^{40}$            \\
OO Aql    & 1.05  & 0.88  &  1.62  &  1.70  & 0.180  &  $10^{39}$               &   $3\times 10^{40}$            \\
V417 Aql  & 1.40  & 0.50  &  1.51  &  1.96  & 0.359  &  $5\times 10^{40}$       &   $3\times 10^{40}$            \\
SS Ari    & 1.31  & 0.40  &  1.54  &  2.07  & 0.372  &  $6\times 10^{40}$       &   $2\times 10^{40}$            \\
V402 Aur  & 1.64  & 0.33  &  1.45  &  2.17  & 0.512  &  $10^{41}$               &   $        10^{40}$            \\
TY Boo    & 0.93  & 0.40  &  1.67  &  2.07  & 0.275  &  $2\times 10^{40}$       &   $2\times 10^{40}$            \\
EF Boo    & 1.61  & 0.82  &  1.46  &  1.73  & 0.282  &  $3\times 10^{40}$       &   $5\times 10^{40}$            \\
AO Cam    & 1.12  & 0.49  &  1.60  &  1.97  & 0.295  &  $3\times 10^{40}$       &   $3\times 10^{40}$            \\
DN Cam    & 1.85  & 0.82  &  1.41  &  1.73  & 0.298  &  $4\times 10^{40}$       &   $5\times 10^{40}$            \\
TX Cnc    & 0.91  & 0.50  &  1.68  &  1.96  & 0.212  &  $9\times 10^{39}$       &   $2\times 10^{40}$            \\
RR Cen    & 2.09  & 0.45  &  1.37  &  2.01  & 0.542  &  $2\times 10^{41}$       &   $2\times 10^{40}$            \\
V752 Cen  & 1.30  & 0.40  &  1.54  &  2.07  & 0.391  &  $6\times 10^{40}$       &   $2\times 10^{40}$            \\
V757 Cen  & 0.88  & 0.59  &  1.70  &  1.88  & 0.212  &  $5\times 10^{39}$       &   $3\times 10^{40}$            \\
VW Cep    & 0.93  & 0.40  &  1.67  &  2.07  & 0.300  &  $2\times 10^{40}$       &   $2\times 10^{40}$            \\
TW Cet    & 1.06  & 0.61  &  1.62  &  1.86  & 0.258  &  $10^{40}$               &   $4\times 10^{40}$            \\
RW Com    & 0.56  & 0.20  &  1.90  &  2.46  & 0.283  &  $10^{40}$               &   $8\times 10^{39}$            \\
RZ Com    & 1.23  & 0.55  &  1.56  &  1.91  & 0.303  &  $3\times 10^{40}$       &   $3\times 10^{40}$            \\
V921 Her  & 2.07  & 0.51  &  1.37  &  1.95  & 0.364  &  $        10^{41}$       &   $2\times 10^{40}$            \\

\hline
\end{tabular*}
\label{tab1}
\end{table}


\begin{thebibliography}{0}
\bibitem{Kopal:1978}
Z.   Kopal,  {\it Close binary systems} (Shapman and Hall LTD, London, 1978).

\bibitem{Shore:1994}
S.N.  Shore,  M.   Livio, and E.P.J.  van den Heuvel,     {\it Interacting binaries} (Springer-Verlag, Berlin-Budapest, 1994).

\bibitem{Hild:2001}
 R.W. Hilditch,     {\it    An introduction to close binary stars}  (Cambridge Univ. Press, Cambridge, 2001).

\bibitem{Boya:2002}
A.A.    Boyarchuk      {\it et al.}, {\it   Mass Transfer in close binary stars} (Teylor and Francis, London, New York, 2002).

\bibitem{Eggleton:2006}
P.P. Eggleton,   {\it   Evolutionary processes in binary and multiple stars} ( Cambridge Univ. Press, Cambridge, 2006).

\bibitem{Khal:2004} K.F. Khaliullin, {\it Dissertation (Russian)} (Sternberg Astronomical Institute, Moscow,
2004).

\bibitem{Vasil:2012}
 B.V.   Vasiliev, {\it Astrophysics and astronomical measurement data} (Fizmatlit, Moscow, 2012);
 Univ. J. Phys. Applic. {\bf 2}, 257 (2014); {\bf 2}, 284 (2014); {\bf 2}, 328 (2014);
 http://astro07.narod.ru.


\bibitem{Cher:2013}
 A.M.  Cherepashchuk, {\it  Close binary stars} (Fizmatlit, Moscow, 2013), vol. I and II.

\bibitem{IJMPE} V.V. Sargsyan, H. Lenske, G.G. Adamian, N.V.   Antonenko, Int. J. Mod. Phys. E {\bf 45},  1850063 (2018).

\bibitem{Adamian:2012}G.G.
 Adamian, N.V.   Antonenko, and  W. Scheid,    Lect. Notes Phys. {\bf 848}, {\it Clusters in Nuclei}
Vol. 2, Ed. by Christian Beck (Springer-Verlag, Berlin, 2012) p. 165.

\bibitem{Adamian:2014}
G.G. Adamian, N.V.   Antonenko, and A.S. Zubov,  Phys. Part. Nucl. {\bf 45}, 848 (2014).





\bibitem{Egg} K. Yakut and P.P. Eggleton,  Astrophys. J.   {\bf 629},  1055 (2005);
K. Gazeas and K. St\c{e}pie\'{n}, MNRAS {\bf 390}, 1577  (2008).

\bibitem{Molnar:2017}
L.A.  Molnar,  D.M.   Van Noord,  K.   Kinemuchi, J.P.  Smolinski,   C.E.  Alexander,   E.M.   Cook,    B.  Jang, H.A.
Kobulnicky, C.J.  Spedden,    and  S.D. Steenwyk, S.D.  arXiv:1704.05502 (2017).

\bibitem{Socia}
Q.J. Socia    {\it et al.},   ApJL {\bf 864},  L32 (2018).

\end{thebibliography}
\end{document}